\documentclass{article}%
\usepackage{graphicx}
\usepackage{amsmath}
\usepackage{amsfonts}
\usepackage{amssymb}%
\begin{document}

\author{Antony Valentini\\Augustus College}

\begin{center}
{\LARGE De Broglie-Bohm Prediction of Quantum Violations for Cosmological
Super-Hubble Modes}

\bigskip

Antony Valentini

\bigskip

\textit{Centre de Physique Th\'{e}orique, Campus de Luminy,}

\textit{Case 907, 13288 Marseille cedex 9, France}

and

\textit{Theoretical Physics Group, Blackett Laboratory, Imperial College
London, Prince Consort Road, London SW7 2AZ, United Kingdom.\footnote{Present
address.}}

email: a.valentini@imperial.ac.uk
\end{center}

The hypothesis of quantum nonequilibrium at the big bang is shown to have
observable consequences. For a scalar field on expanding space, we show that
relaxation to quantum equilibrium (in de Broglie-Bohm theory) is suppressed
for field modes whose quantum time evolution satisfies a certain inequality,
resulting in a `freezing' of early quantum nonequilibrium for these particular
modes. For an early radiation-dominated expansion, the inequality implies a
corresponding physical wavelength that is larger than the (instantaneous)
Hubble radius. These results make it possible, for the first time, to make
quantitative predictions for nonequilibrium deviations from quantum theory, in
the context of specific cosmological models. We discuss some possible
consequences: corrections to inflationary predictions for the cosmic microwave
background, non-inflationary super-Hubble field correlations, and relic
nonequilibrium particles.

\ 

1 Introduction

2 De Broglie-Bohm Scalar Field on Expanding Space

3 Preliminary Discussion for a Decoupled Mode

\qquad{\small 3.1 Relaxation for Sub-Hubble Modes in the Minkowski Limit}

\qquad{\small 3.2 Freezing of the Wave Function for Super-Hubble Modes}

4 Inequality for the Freezing of Quantum Nonequilibrium

5 Generalisations

\qquad{\small 5.1 Mixed States}

\qquad{\small 5.2 Interacting Fields}

6 General Implications of the Freezing Inequality

7 Relaxation for Modes Violating the Freezing Inequality

8 Time Evolution of $\left\langle \hat{n}_{\mathbf{k}r}\right\rangle $ for a
Free Field

9 Approximate Solutions for $\left\langle \hat{n}_{\mathbf{k}r}\right\rangle
_{t}$ Satisfying the Freezing Inequality

10 Possible Consequences of Early Nonequilibrium Freezing

\qquad{\small 10.1 Corrections to Inflationary Predictions for the CMB}

\qquad{\small 10.2 Super-Hubble Correlations without Inflation?}

\qquad{\small 10.3 Relic Nonequilibrium Particles}

11 Conclusion

\bigskip

\bigskip

\bigskip

\bigskip

\bigskip

\bigskip

\bigskip

\bigskip

\bigskip

\bigskip

\bigskip

\bigskip

\bigskip

\bigskip

\bigskip

\bigskip\bigskip

\bigskip

\bigskip

\section{Introduction}

Hidden-variables theories, such as the pilot-wave theory of de Broglie
\cite{deB28,BV} and Bohm \cite{B52}, reproduce quantum theory for a particular
`equilibrium' distribution of hidden parameters. But allowing arbitrary
distributions (analogous to non-thermal distributions in classical physics)
opens up the possibility of new, `nonequilibrium' physics that lies outside
the domain of quantum physics
\cite{AV91a,AV91b,AV92,AV96,AV01,AV02a,AV02b,AVPr02,AVSig,AVBHs,PV06,AV07}.
Such new physics may have existed in the very early universe, with relaxation
to quantum equilibrium having taken place during the violence of the big bang
\cite{AV91a,AV91b,AV92,AV96,AV01,AV07}. In this paper, the hypothesis of early
quantum nonequilibrium is shown to have observable consequences today.

The concept of quantum nonequilibrium has been discussed for general
(deterministic)\ hidden-variables theories \cite{AV02a,AV02b,AVSig,PV06}. For
the specific case of de Broglie-Bohm theory, it amounts to having
configurations with a distribution $P$ that differs from the usual Born-rule
distribution $\left\vert \Psi\right\vert ^{2}$ (for a pure subensemble with
wave function $\Psi$) \cite{AV91a,AV91b,AV92,PV06}. There were several
motivations for proposing that the early universe began in a state of quantum
nonequilibrium. Let us briefly summarise them.

There seems to be a peculiar `conspiracy' at the heart of modern physics,
whereby quantum nonlocality cannot be used to send practical instantaneous
signals. In hidden-variables theories, this conspiracy is explained as a
contingency of the quantum equilibrium state. Nonlocal signalling is generally
possible out of equilibrium (suggesting the existence of an underlying
preferred foliation of spacetime \cite{AVSim}); whereas in equilibrium,
nonlocal effects cancel out at the statistical level
\cite{AV91b,AV92,AV02a,AV02b}. Our inability to convert entanglement into
practical nonlocal signals is then not a law of physics, but a contingency of
the equilibrium state. Similarly, standard uncertainty-principle limitations
on measurements are also contingencies of equilibrium \cite{AV91b,AV92,AVPr02}%
. There is a parallel here with the classical thermodynamic heat death: in the
complete absence of temperature differences, it would be impossible to convert
heat into work, and yet such a limitation would be a mere contingency of the
state, and not a law of physics.

Furthermore, it has been shown that relaxation towards quantum equilibrium
occurs, in pilot-wave dynamics, in similar fashion to thermal relaxation in
classical dynamics (under analogous conditions and with similar caveats)
\cite{AV91a,AV92,AV01,VW05,Sky}. Given that all physical systems to which we
have access have undergone a long and violent astrophysical history, it is
then possible to understand the ubiquitous quantum noise we see around us as,
in effect, a remnant of the big bang.

On this view, the effectively local and indeterministic quantum physics we
experience today emerged via relaxation processes (presumably occurring close
to the big bang) out of a fundamentally nonlocal and deterministic physics ---
a physics whose details are currently screened off from view, by the
all-pervading statistical noise. For as equilibrium is approached, the
possibility of instantaneous signalling disappears, and statistical
uncertainty emerges. In effect, a hidden-variables analogue of the classical
heat death has actually occurred in our universe, explaining the above `conspiracy'.

The assumption of early quantum nonequilibrium was also proposed as a possible
alternative resolution of the cosmological horizon problem (which persists
even in some inflationary models \cite{VT00}): the resulting early nonlocality
might explain the otherwise puzzling homogeneity of the universe at early
times \cite{AV91b,AV92,AV96,AV02b}.

The search for early quantum nonequilibrium may also be motivated simply on
the grounds that de Broglie-Bohm theory (and indeed any deterministic
hidden-variables theory) certainly \textit{allows} nonequilibrium to occur. We
have an alternative formulation of quantum physics, which yields standard
quantum theory in the equilibrium limit, and which yields departures from
standard quantum theory outside that limit. It seems natural to explore this
possible new physics, and in particular to test for it experimentally, as far
as one can. If nothing else, setting experimental bounds on the existence of
quantum nonequilibrium can provide new bounds on possible deviations from
quantum theory \cite{AV07}.

Finally, if hidden-variables theories are taken seriously, one is obliged to
take the possibility of nonequilibrium seriously as well: for it is only in
nonequilibrium that the underlying details become visible. If the world were
always and everywhere in quantum equilibrium, the details of de Broglie-Bohm
trajectories (for example) would be forever shielded from experimental test.
De Broglie-Bohm theory as a whole would then be unacceptable as a scientific
theory. And much the same could be said for hidden-variables theories in general.

Given the above motivations, the idea that the universe relaxed to quantum
equilibrium from an earlier nonequilibrium state is plausible enough. However,
to be a scientific theory it is essential to make new, quantitative
predictions. The new physics of systems in quantum nonequilibrium has been
explored in some detail \cite{AV91b,AV92,AV01,AV02a,AVPr02,AVBHs,PV06,AV07},
and a specific signature of nonequilibrium has been developed
\cite{AVSig,PV06}. It has also been shown that if nonequilibrium were present
at the beginning of an inflationary phase, then there would be observable
consequences for the statistics of the temperature anisotropies imprinted on
the cosmic microwave background (CMB) \cite{AV07,AV08}. Further, heuristic
arguments have been given, suggesting that relaxation might be suppressed for
long-wavelength field modes on expanding space \cite{AV07} (a suggestion that
forms the starting point for the present work); and that, if relic
cosmological particles decoupled sufficiently early, they might still be in
nonequilibrium today \cite{AV01,AV07}. However, so far, no definite
quantitative predictions have been made. The aim of this paper is to fill this gap.

For the first time, given a specific cosmological model, we are able to point
to precisely where quantum nonequilibrium could be found. We accomplish this
by studying the evolution of nonequilibrium distributions for a scalar field
on expanding space. We show that relaxation is suppressed for field modes
whose quantum time evolution satisfies a certain inequality. For these
particular modes, early quantum nonequilibrium is `frozen'. For a
radiation-dominated expansion, the inequality implies a physical wavelength
larger than the (instantaneous) Hubble radius. On the basis of these results,
it is possible to make quantitative predictions for nonequilibrium deviations
from quantum theory, in the context of a given cosmological model. As we shall
see, there are a number of possible consequences: in particular, infra-red
corrections to inflationary predictions for the CMB, and relic nonequilibrium
particles at low energies.

\section{De Broglie-Bohm Scalar Field on Expanding Space}

For simplicity we consider a flat metric,%
\begin{equation}
d\tau^{2}=dt^{2}-a^{2}d\mathbf{x}^{2}\ ,\label{metric}%
\end{equation}
where $a(t)$ is the scale factor, $H\equiv\dot{a}/a$ is the Hubble parameter,
and $H^{-1}$ is the Hubble radius. As is customary, we take $a_{0}=1$ today
(at time $t_{0}$), so that $|d\mathbf{x}|$ is a comoving distance (or proper
distance today). At time $t$, field modes have physical wavelengths
$\lambda_{\mathrm{phys}}=a(t)\lambda$, where $\lambda=2\pi/k$ is a comoving
wavelength (or proper wavelength today) and $k=|\mathbf{k}|$ is the comoving
wave number.

We consider a free (minimally-coupled) massless scalar field $\phi$ with a
Lagrangian density $\mathcal{L}=\frac{1}{2}g^{1/2}\partial_{\alpha}%
\phi\partial^{\alpha}\phi$ or%
\begin{equation}
\mathcal{L}=\tfrac{1}{2}a^{3}\dot{\phi}^{2}-\tfrac{1}{2}a(\mathbf{\nabla}%
\phi)^{2}\ . \label{Lagden}%
\end{equation}
The action is $\int dt\int d^{3}\mathbf{x}\;\mathcal{L}$. We then have a
canonical momentum density $\pi=\partial\mathcal{L}/\partial\dot{\phi}%
=a^{3}\dot{\phi}$ and a Hamiltonian density%
\begin{equation}
\mathcal{H}=\tfrac{1}{2}\frac{\pi^{2}}{a^{3}}+\tfrac{1}{2}a(\mathbf{\nabla
}\phi)^{2}\ . \label{Hamden}%
\end{equation}
Here, it is convenient to write the dynamics in Fourier space. Expressing
$\phi(\mathbf{x})$ in terms of its Fourier components%
\[
\phi_{\mathbf{k}}=\frac{1}{(2\pi)^{3/2}}\int d^{3}\mathbf{x}\;\phi
(\mathbf{x})e^{-i\mathbf{k}\cdot\mathbf{x}}\ ,
\]
and writing%
\[
\phi_{\mathbf{k}}=\frac{\sqrt{V}}{(2\pi)^{3/2}}\left(  q_{\mathbf{k}%
1}+iq_{\mathbf{k}2}\right)
\]
for real $q_{\mathbf{k}r}$ ($r=1$, $2$), where $V$ is a box normalisation
volume, the Lagrangian $L=\int d^{3}\mathbf{x}\;\mathcal{L}$ becomes%
\[
L=\sum_{\mathbf{k}r}\frac{1}{2}\left(  a^{3}\dot{q}_{\mathbf{k}r}^{2}%
-ak^{2}q_{\mathbf{k}r}^{2}\right)  \ .
\]
(For $V\rightarrow\infty$, $\frac{1}{V}\sum_{\mathbf{k}}\rightarrow\frac
{1}{(2\pi)^{3}}\int d^{3}\mathbf{k}$ and $V\delta_{\mathbf{kk%
\acute{}%
}}\rightarrow(2\pi)^{3}\delta^{3}(\mathbf{k}-\mathbf{k}%
\acute{}%
)$. Since $\phi$ is real, we have $\phi_{\mathbf{k}}^{\ast}=\phi_{-\mathbf{k}%
}$ or $q_{\mathbf{k}1}=q_{-\mathbf{k}1}$, $q_{\mathbf{k}2}=-q_{-\mathbf{k}2}$.
A sum over physical degrees of freedom should be restricted to half the values
of $\mathbf{k}$, for example $k_{z}>0$.) Introducing the canonical momenta%
\[
\pi_{\mathbf{k}r}\equiv\frac{\partial L}{\partial\dot{q}_{\mathbf{k}r}}%
=a^{3}\dot{q}_{\mathbf{k}r}\ ,
\]
the Hamiltonian $H=\int d^{3}\mathbf{x}\;\mathcal{H}$ becomes%
\[
H=\sum_{\mathbf{k}r}H_{\mathbf{k}r}\ ,
\]
with%
\[
H_{\mathbf{k}r}=\frac{1}{2a^{3}}\pi_{\mathbf{k}r}^{2}+\frac{1}{2}%
ak^{2}q_{\mathbf{k}r}^{2}\ .
\]

Pilot-wave field theory is defined in terms of the functional Schr\"{o}dinger
picture, with a preferred foliation of spacetime
\cite{B52,AV92,AV96,BH84,BHK87,Holl93,BandH,Kal94,StruyPR}. Here, the
Schr\"{o}dinger equation for $\Psi=\Psi\lbrack q_{\mathbf{k}r},t]$ is%
\begin{equation}
i\frac{\partial\Psi}{\partial t}=\sum_{\mathbf{k}r}\left(  -\frac{1}{2a^{3}%
}\frac{\partial^{2}}{\partial q_{\mathbf{k}r}^{2}}+\frac{1}{2}ak^{2}%
q_{\mathbf{k}r}^{2}\right)  \Psi\ ,\label{Sch2'}%
\end{equation}
which implies the continuity equation%
\begin{equation}
\frac{\partial\left\vert \Psi\right\vert ^{2}}{\partial t}+\sum_{\mathbf{k}%
r}\frac{\partial}{\partial q_{\mathbf{k}r}}\left(  \left\vert \Psi\right\vert
^{2}\frac{1}{a^{3}}\frac{\partial S}{\partial q_{\mathbf{k}r}}\right)
=0\label{Cont2}%
\end{equation}
and the de Broglie velocities%
\begin{equation}
\frac{dq_{\mathbf{k}r}}{dt}=\frac{1}{a^{3}}\frac{\partial S}{\partial
q_{\mathbf{k}r}}\label{deB2}%
\end{equation}
(where $\Psi=\left\vert \Psi\right\vert e^{iS}$). The `pilot wave' $\Psi$ is
interpreted as a physical field in configuration space, guiding the time
evolution of an individual field $\phi(\mathbf{x},t)$ in 3-space. (Note that a
similar construction may be given in any globally-hyperbolic spacetime, by
choosing a preferred foliation \cite{AVBHs}, so there is no need for spatial homogeneity.)

Over an ensemble of field configurations guided by the same pilot wave $\Psi$,
there will be some (in principle arbitrary) initial distribution
$P[q_{\mathbf{k}r},t_{i}]$, whose time evolution $P[q_{\mathbf{k}r},t]$ will
be determined by%
\begin{equation}
\frac{\partial P}{\partial t}+\sum_{\mathbf{k}r}\frac{\partial}{\partial
q_{\mathbf{k}r}}\left(  P\frac{1}{a^{3}}\frac{\partial S}{\partial
q_{\mathbf{k}r}}\right)  =0\ . \label{ContP1}%
\end{equation}
If $P[q_{\mathbf{k}r},t_{i}]=\left\vert \Psi\lbrack q_{\mathbf{k}r}%
,t_{i}]\right\vert ^{2}$, then $P[q_{\mathbf{k}r},t]=\left\vert \Psi\lbrack
q_{\mathbf{k}r},t]\right\vert ^{2}$ for all $t$, and one obtains empirical
agreement with standard quantum field theory
\cite{B52,BHK87,Holl93,BandH,StruyPR,Kal94}. On the other hand, for an initial
nonequilibrium distribution $P[q_{\mathbf{k}r},t_{i}]\neq\left\vert
\Psi\lbrack q_{\mathbf{k}r},t_{i}]\right\vert ^{2}$, for as long as $P$
remains in nonequilibrium the predicted statistics will generally differ from
those of quantum field theory. In any case, whatever the distribution $P$ may
be (equilibrium or nonequilibrium), its time evolution will be given by
(\ref{ContP1}).

\section{Preliminary Discussion for a Decoupled Mode}

A proper treatment of nonequilibrium freezing is given in sections 4 and 5. As
we shall see, our treatment is applicable to arbitrary (entangled, mixed, and
interacting) quantum states. As a preliminary exercise, in this section we
shall discuss some elementary features for the simple case of a single
decoupled mode $\mathbf{k}$ of a free field in a pure quantum state.

From equations (\ref{Sch2'}), (\ref{deB2}), and writing $\Psi=\psi
_{\mathbf{k}}(q_{\mathbf{k}1},q_{\mathbf{k}2},t)\varkappa$ where $\varkappa$
depends only on degrees of freedom for modes $\mathbf{k}%
\acute{}%
\neq\mathbf{k}$, we find that the wave function $\psi_{\mathbf{k}}$ of a
decoupled mode $\mathbf{k}$ satisfies%
\begin{equation}
i\frac{\partial\psi_{\mathbf{k}}}{\partial t}=-\frac{1}{2a^{3}}\left(
\frac{\partial^{2}}{\partial q_{\mathbf{k}1}^{2}}+\frac{\partial^{2}}{\partial
q_{\mathbf{k}2}^{2}}\right)  \psi_{\mathbf{k}}+\frac{1}{2}ak^{2}\left(
q_{\mathbf{k}1}^{2}+q_{\mathbf{k}2}^{2}\right)  \psi_{\mathbf{k}}\ ,
\label{S2D}%
\end{equation}
while the de Broglie velocities for the mode amplitudes $(q_{\mathbf{k}%
1},q_{\mathbf{k}2})$ are%
\begin{equation}
\dot{q}_{\mathbf{k}1}=\frac{1}{a^{3}}\frac{\partial s_{\mathbf{k}}}{\partial
q_{\mathbf{k}1}},\ \ \ \ \dot{q}_{\mathbf{k}2}=\frac{1}{a^{3}}\frac{\partial
s_{\mathbf{k}}}{\partial q_{\mathbf{k}2}} \label{deB2D}%
\end{equation}
(with $\psi_{\mathbf{k}}=\left\vert \psi_{\mathbf{k}}\right\vert
e^{is_{\mathbf{k}}}$). The time evolution of the marginal distribution
$\rho_{\mathbf{k}}(q_{\mathbf{k}1},q_{\mathbf{k}2},t)$ will then be given by%
\begin{equation}
\frac{\partial\rho_{\mathbf{k}}}{\partial t}+\sum_{r=1,\ 2}\frac{\partial
}{\partial q_{\mathbf{k}r}}\left(  \rho_{\mathbf{k}}\frac{1}{a^{3}}%
\frac{\partial s_{\mathbf{k}}}{\partial q_{\mathbf{k}r}}\right)  =0\ .
\label{C2D}%
\end{equation}

Equations (\ref{S2D})--(\ref{C2D}) are identical to those of pilot-wave
dynamics for an ensemble of nonrelativistic particles of time-dependent `mass'
$m=a^{3}$ moving in the $q_{\mathbf{k}1}-q_{\mathbf{k}2}$ plane in a harmonic
oscillator potential with time-dependent angular frequency $\omega=k/a$. We
may then discuss relaxation (and relaxation suppression) for a decoupled field
mode in terms of relaxation (and relaxation suppression) for a nonrelativistic
two-dimensional harmonic oscillator.

Before doing so, let us recall what is already known about relaxation in
pilot-wave dynamics.

For a system with configuration $q$ and wave function $\psi$, the $H$-function%
\begin{equation}
H=\int dq\;\rho\ln(\rho/\left\vert \psi\right\vert ^{2})
\end{equation}
(the relative negentropy of an arbitrary distribution $\rho$ with respect to
$\left\vert \psi\right\vert ^{2}$) obeys a coarse-graining $H$-theorem similar
to the classical one \cite{AV91a,AV92,AV01}. Introducing a coarse-graining in
configuration space, and assuming appropriate initial conditions for $\rho$
and $\psi$, the coarse-grained function $\bar{H}(t)$ will begin to decrease
with time, corresponding to an evolution of the coarse-grained density
$\bar{\rho}$ towards $\overline{\left\vert \psi\right\vert ^{2}}$. This
`subquantum $H$-theorem' formalises a simple intuitive idea: because $\rho$
and $\left\vert \psi\right\vert ^{2}$ obey the same continuity equation, they
behave like two classical fluids that are `stirred' by the same velocity
field, thereby tending to become indistinguishable on a coarse-grained level.

Such relaxation has been studied numerically, on a static spacetime
background, for simple one- and two-dimensional systems
\cite{AV92,AV01,VW05,Sky}. For an ensemble of nonrelativistic particles in a
two-dimensional box, with a wave function consisting of a superposition of the
first 16 modes, it was found that relaxation occurs very efficiently, with an
approximately exponential decay $\bar{H}(t)\approx\bar{H}_{0}%
e^{-t/t_{\mathrm{c}}}$ of the coarse-grained $H$-function (over a timescale
$t_{\mathrm{c}}$) \cite{VW05}. Similar results have been obtained for an
ensemble of nonrelativistic particles in a two-dimensional harmonic oscillator
potential \cite{Sky}. As discussed in ref. \cite{VW05}, the numerical
timescale $t_{\mathrm{c}}$ was found to be in approximate agreement with a
theoretical relaxation timescale $\tau$ defined by $1/\tau^{2}\equiv
-(1/\bar{H})d^{2}\bar{H}/dt^{2}$ \cite{AV92}. For a particle of mass $m$, and
using a sufficiently small coarse-graining length $\varepsilon$, a rough
order-of-magnitude estimate yields $\tau\sim1/\left(  \varepsilon
m^{1/2}(\Delta E)^{3/2}\right)  $, where $\Delta E$ is the quantum energy
spread associated with $\psi$ \cite{AV01,VW05}. (The quantity $\tau$ is
analogous to the scattering time of classical kinetic theory: one expects a
significant approach to equilibrium over timescales of order $\tau$.) If we
choose a `natural' value $\varepsilon\sim1/\Delta p$, where $\Delta p$ is the
quantum momentum spread, then taking $\Delta E\sim(\Delta p)^{2}/2m$ one has
the simple (and rough) result%
\begin{equation}
\tau\sim\Delta t\equiv1/\Delta E\ , \label{tau}%
\end{equation}
where $\Delta t$ is the quantum timescale over which the wave function $\psi$ evolves.

\subsection{Relaxation for Sub-Hubble Modes in the Minkowski Limit}

One expects that in the short-wavelength limit, $\lambda_{\mathrm{phys}%
}<<H^{-1}$, the above equations (\ref{S2D})--(\ref{C2D}) will reduce to those
for a decoupled mode $\mathbf{k}$ on Minkowski spacetime, because (roughly
speaking) the timescale $\Delta t\propto\lambda_{\mathrm{phys}}$ over which
$\psi_{\mathbf{k}}=\psi_{\mathbf{k}}(q_{\mathbf{k}1},q_{\mathbf{k}2},t)$
evolves will be much smaller than the expansion timescale $H^{-1}\equiv
a/\dot{a}$ \cite{AV07}.

To obtain a more precise and rigorous statement, note first that at any time
$t$ the Hamiltonian $\hat{H}(t)$ appearing in the Schr\"{o}dinger equation
(\ref{S2D}) has the same eigenfunctions and eigenvalues as are usually
obtained for a two-dimensional harmonic oscillator of (instantaneous) mass
$m=a^{3}$ and angular frequency $\omega=k/a$. Thus, for quantum numbers
$n_{1}$, $n_{2}=0,1,2,$ ... , we have energy eigenfunctions $\phi_{n_{1}%
}(q_{\mathbf{k}1},t)\phi_{n_{2}}(q_{\mathbf{k}2},t)$ and eigenvalues
$E_{\mathbf{k}}(t)=(1+n_{1}+n_{2})\omega(t)$. (The time dependence in
$\phi_{n_{1}}(q_{\mathbf{k}1},t)$ and $\phi_{n_{2}}(q_{\mathbf{k}2},t)$ comes,
of course, from the time dependence of $m=a^{3}$ and $\omega=k/a$.) The wave
function at any time $t$ may then be expanded in terms of these energy
eigenstates,%
\[
\psi_{\mathbf{k}}(q_{\mathbf{k}1},q_{\mathbf{k}2},t)=\sum_{n_{1},\ n_{2}%
}c_{n_{1},n_{2}}(t)\phi_{n_{1}}(q_{\mathbf{k}1},t)\phi_{n_{2}}(q_{\mathbf{k}%
2},t)\ ,
\]
and the quantum energy spread $\Delta E_{\mathbf{k}}\equiv\sqrt{\left\langle
E_{\mathbf{k}}^{2}\right\rangle -\left\langle E_{\mathbf{k}}\right\rangle
^{2}}$ will be%
\[
\Delta E_{\mathbf{k}}=\Delta n_{\mathbf{k}}\cdot\omega\ ,
\]
where $n_{\mathbf{k}}\equiv n_{1}+n_{2}$. If we consider a subsequent
evolution over a time $\delta t<<H^{-1}$, where $H^{-1}$ is the timescale over
which the Hamiltonian $\hat{H}(t)$ changes, then the Hamiltonian (together
with its eigenfunctions and eigenvalues) will be almost constant during
$(t,t+\delta t)$, and in this interval the wave function $\psi_{\mathbf{k}}$
will evolve like that of a conventional two-dimensional oscillator, with an
evolution timescale%
\[
\Delta t\equiv\frac{1}{\Delta E_{\mathbf{k}}}=\frac{1}{\Delta n_{\mathbf{k}}%
}\frac{1}{\omega}=\frac{1}{\Delta n_{\mathbf{k}}}\frac{\lambda_{\mathrm{phys}%
}}{2\pi}%
\]
(where we have $\hbar=1$). Significant evolution of $\psi_{\mathbf{k}}$ over
the interval $(t,t+\delta t)$ can occur only if $\Delta t<<H^{-1}$ or%
\begin{equation}
\lambda_{\mathrm{phys}}<<\Delta n_{\mathbf{k}}\cdot H^{-1}\ . \label{Min}%
\end{equation}

We may then take (\ref{Min}) to be a good characterisation of the
short-wavelength or Minkowski limit. In this limit, over timescales $\Delta
t\equiv1/\Delta E_{\mathbf{k}}<<H^{-1}$, the wave function $\psi_{\mathbf{k}}$
evolves just as it would on Minkowski spacetime. On such timescales, the scale
factor $a$ is approximately constant, and the equations (\ref{S2D}%
)--(\ref{C2D}) reduce to those of pilot-wave dynamics for an ensemble of
nonrelativistic particles of constant mass $m=a^{3}$ moving in a
two-dimensional harmonic oscillator potential of constant angular frequency
$\omega=k/a$. From the numerical results for the latter case \cite{Sky} we may
deduce that, in the Minkowski limit, for a decoupled mode $\mathbf{k}$ in a
superposition $\left\vert \psi_{\mathbf{k}}\right\rangle \sim\left\vert
1_{\mathbf{k}}\right\rangle +\left\vert 2_{\mathbf{k}}\right\rangle
+\left\vert 3_{\mathbf{k}}\right\rangle +\ ...$ of many different states of
definite occupation number, the distribution $\rho_{\mathbf{k}}(q_{\mathbf{k}%
1},q_{\mathbf{k}2},t)$ of the mode amplitudes will relax to equilibrium,
$\rho_{\mathbf{k}}\rightarrow\left\vert \psi_{\mathbf{k}}\right\vert ^{2}$ (on
a coarse-grained level, again assuming appropriate initial conditions), on a
timescale $\tau$ given roughly by (\ref{tau}) or%
\[
\tau\sim\frac{1}{\Delta E_{\mathbf{k}}}=\frac{1}{\Delta n_{\mathbf{k}}}%
\frac{1}{\omega}\ .
\]

\subsection{Freezing of the Wave Function for Super-Hubble Modes}

In contrast, in the long-wavelength limit,%
\begin{equation}
\lambda_{\mathrm{phys}}>>\Delta n_{\mathbf{k}}\cdot H^{-1}\ , \label{long}%
\end{equation}
we have $\Delta t\equiv1/\Delta E_{\mathbf{k}}>>H^{-1}$ and the change in the
Hamiltonian $\hat{H}(t)$ over timescales $H^{-1}$ may be treated as a sudden
perturbation, leading to the conclusion that the wave function $\psi
_{\mathbf{k}}$ is approximately static --- or `frozen' --- over timescales
$H^{-1}$.

More precisely, let us again consider an evolution over an interval
$(t,t+\delta t)$ --- but now with $\delta t$ of order $H^{-1}$, so that the
Hamiltonian $\hat{H}(t)$ changes significantly. We may write $\hat{H}(t+\delta
t)=\hat{H}(t)+\delta\hat{H}$, where $\delta\hat{H}$ is comparable to $\hat
{H}(t)$. In the limit $\lambda_{\mathrm{phys}}>>\Delta n_{\mathbf{k}}\cdot
H^{-1}$, the timescale $\Delta t\equiv1/\Delta E_{\mathbf{k}}$ associated with
the `unperturbed' Hamiltonian $\hat{H}(t)$ will be large compared to the
timescale $H^{-1}$ over which the Hamiltonian changes. We may then treat the
change $\delta\hat{H}$ as a sudden perturbation, applied over a timescale that
is short compared to the natural timescale of the system. By standard
reasoning (for example, ref. \cite{Shank}), we deduce that $\psi_{\mathbf{k}}$
hardly changes over the interval $(t,t+\delta t)$, that is, that
$\psi_{\mathbf{k}}$ is essentially static over timescales $H^{-1}$.

Note that the above freezing of the wave function on timescales $H^{-1}$ need
not occur for \textit{all} super-Hubble modes, since for any $\lambda
_{\mathrm{phys}}>H^{-1}$ the long-wavelength condition (\ref{long}) will be
violated if $\Delta n_{\mathbf{k}}$ is sufficiently large. On the other hand,
of course, for any given value of $\Delta n_{\mathbf{k}}$, the condition
(\ref{long}) will be satisfied for sufficiently large $\lambda_{\mathrm{phys}%
}$ and the wave function will indeed be frozen.

If $\psi_{\mathbf{k}}$ is frozen over timescales $H^{-1}$, then the
equilibrium density $\left\vert \psi_{\mathbf{k}}\right\vert ^{2}$ is also
frozen over timescales $H^{-1}$. Because the evolution of $\left\vert
\psi_{\mathbf{k}}(q_{\mathbf{k}1},q_{\mathbf{k}2},t)\right\vert ^{2}$ is
driven by the de Broglie velocity field $(\dot{q}_{\mathbf{k}1},\dot
{q}_{\mathbf{k}2})$, in accordance with the continuity equation%
\begin{equation}
\frac{\partial\left\vert \psi_{\mathbf{k}}\right\vert ^{2}}{\partial t}%
+\frac{\partial}{\partial q_{\mathbf{k}1}}\left(  \left\vert \psi_{\mathbf{k}%
}\right\vert ^{2}\dot{q}_{\mathbf{k}1}\right)  +\frac{\partial}{\partial
q_{\mathbf{k}2}}\left(  \left\vert \psi_{\mathbf{k}}\right\vert ^{2}\dot
{q}_{\mathbf{k}2}\right)  =0\ , \label{ce}%
\end{equation}
we then expect that the trajectories $(q_{\mathbf{k}1}(t),q_{\mathbf{k}2}(t))$
will also be frozen over timescales $H^{-1}$. (In principle, of course,
(\ref{ce}) can have solutions with an essentially static density $\left\vert
\psi_{\mathbf{k}}\right\vert ^{2}$ and a non-negligible velocity field
$(\dot{q}_{\mathbf{k}1},\dot{q}_{\mathbf{k}2})$, but we expect these to occur
only in exceptional circumstances. And in any case, because the phase gradient
$\partial s_{\mathbf{k}}/\partial q_{\mathbf{k}r}$ is also frozen over
timescales $H^{-1}$, from (\ref{deB2D}) we see that the velocities $\dot
{q}_{\mathbf{k}r}$ become smaller as the scale factor $a$ increases over
expansion timescales $H^{-1}$.) Assuming this to be the case, it then follows
that an arbitrary nonequilibrium distribution $\rho_{\mathbf{k}}\neq\left\vert
\psi_{\mathbf{k}}\right\vert ^{2}$, evolving in time according to the same
continuity equation (\ref{ce}), will also be frozen over timescales $H^{-1}$.
In other words, at least in this simple case of a decoupled field mode,
initial quantum nonequilibrium will be frozen on timescales of order the
expansion timescale $H^{-1}$. (This is reminiscent of the well-known
`freezing' of super-Hubble modes in the theory of cosmological perturbations
\cite{Pad93,LL00}.)

The above reasoning then suggests a mechanism, whereby the rapid expansion of
space at early times can suppress the normal process of relaxation to quantum
equilibrium, raising the possibility that remnants of early nonequilibrium
could have survived to the present day \cite{AV01,AV07}. However, our
treatment so far is rather limited. We have considered only a free, decoupled
mode in a pure quantum state. It is only expected, and not generally proven,
that a frozen $\left\vert \psi_{\mathbf{k}}\right\vert ^{2}$ will be
associated with a family of frozen trajectories. And, perhaps most seriously,
while it seems significant to demonstrate nonequilibrium freezing over the
(time-dependent) expansion timescale $H^{-1}$, in a standard --- say
radiation-dominated --- expansion we have $H^{-1}\rightarrow0$ as
$t\rightarrow0$, so by itself nonequilibrium freezing over the timescale
$H^{-1}$ does not tell us very much about the possible survival of initial
nonequilibrium. These limitations will be overcome in the following two
sections. We shall first derive a rigorous condition for nonequilibrium
freezing, applicable to an arbitrary time interval and to any (generally
entangled) pure quantum state of a free field. Then, we shall generalise this
condition to mixed states and to interacting fields.

\section{Inequality for the Freezing of Quantum Nonequilibrium}

To study nonequilibrium freezing over arbitrary time intervals and for
arbitrary quantum states, we shall examine the behaviour of the trajectories
themselves (instead of the behaviour of their guiding wave functions), thereby
obtaining a direct constraint on the evolution of nonequilibrium distributions.

Mathematically, as we saw in section 2, the field system is equivalent to a
collection of non-interacting one-dimensional harmonic oscillators with
positions $q_{\mathbf{k}r}$ (and with time-dependent masses $m=a^{3}$ and
time-dependent angular frequencies $\omega=k/a$). The Hamiltonian operator is
$\hat{H}=\sum_{\mathbf{k}r}\hat{H}_{\mathbf{k}r}$, with%
\[
\hat{H}_{\mathbf{k}r}=\frac{\hat{\pi}_{\mathbf{k}r}^{2}}{2a^{3}}+\frac{1}%
{2}a^{3}\omega^{2}\hat{q}_{\mathbf{k}r}^{2}\ .
\]
Each $\hat{H}_{\mathbf{k}r}$ has (time-dependent) energy eigenvalues
$E_{\mathbf{k}r}=(n_{\mathbf{k}r}+\frac{1}{2})\omega$, where $n_{\mathbf{k}%
r}=0,1,2,....$ . (Because of the explicit time dependence in the Hamiltonian,
the mean energy is of course not conserved: $d\left\langle \hat{H}%
\right\rangle /dt=\left\langle \partial\hat{H}/\partial t\right\rangle \neq
0$.) For an arbitrary wave functional $\Psi\lbrack q_{\mathbf{k}r},t]$, the de
Broglie velocity field is given by (\ref{deB2}), and the evolution of an
arbitrary ensemble distribution $P[q_{\mathbf{k}r},t]$ will be driven by this
velocity field via the continuity equation (\ref{ContP1}).

Note that the use of a classical spacetime background must break down in the
limit $t\rightarrow0$. The equations defining our model can be trusted only
down to some minimum initial time $t_{i}$. For example, very optimistically,
one might take the `initial time' to be of order the Planck time, $t_{i}\sim
t_{\mathrm{P}}\sim10^{-43}\ \mathrm{s}$.

Now, an initial nonequilibrium distribution $P[q_{\mathbf{k}r},t_{i}]\neq
|\Psi\lbrack q_{\mathbf{k}r},t_{i}]|^{2}$ can in general relax to equilibrium
(on a coarse-grained level) only if the trajectories wander sufficiently far
over the region of configuration space where $|\Psi|^{2}$ is concentrated;
otherwise, for example, if $P$ were initially small in regions where
$|\Psi|^{2}$ is large, $P$ could remain so, and equilibrium would never be
reached. We may then write a simple condition for initial nonequilibrium to be
`frozen', by considering the displacements of the trajectories, and requiring
that the (equilibrium) mean magnitude of the displacements be smaller than the
width of the wave packet.

Let us write the total configuration of the system as $q(t)$. Note that
$\Psi\lbrack q,t]$ is in general an entangled function of all the
$q_{\mathbf{k}r}$'s. Even so, given the initial distributions $P[q,t_{i}]$ and
$|\Psi\lbrack q,t_{i}]|^{2}$, one may calculate the corresponding marginals
for just one $q_{\mathbf{k}r}$ (for some given $\mathbf{k}r$). If the
resulting two marginals are equal or unequal, we may say that we have
equilibrium or nonequilibrium respectively, for the given degree of freedom
$q_{\mathbf{k}r}$. In this sense, it is clearly possible for some of the
$q_{\mathbf{k}r}$'s to be in nonequilibrium while the others are in equilibrium.

Let us now consider the motion $q_{\mathbf{k}r}(t)$ of one degree of freedom,
for some given $\mathbf{k}r$, over a time interval $[t_{i},t_{f}]$. An initial
point $q_{\mathbf{k}r}(t_{i})$ undergoes a final displacement $\delta
q_{\mathbf{k}r}(t_{f})=\int_{t_{i}}^{t_{f}}dt\ \dot{q}_{\mathbf{k}r}(t)$,
where the velocity $\dot{q}_{\mathbf{k}r}$ is given by (\ref{deB2}%
).\footnote{Note that trajectories in one-dimensional $q_{\mathbf{k}r}$-space
do move past each other, being components of higher-dimensional trajectories
$q(t)$ (unlike in a strictly one-dimensional system, where the
single-valuedness of the velocity field prevents trajectories from crossing).}
Let $\Delta q_{\mathbf{k}r}(t)$ be the width --- with respect to
$q_{\mathbf{k}r}$ --- of the quantum distribution $|\Psi\lbrack q,t]|^{2}$ at
time $t$. If the whole family of trajectories $q_{\mathbf{k}r}(t)$ (with fixed
$\mathbf{k}r$ and arbitrary initial total configurations $q(t_{i})$) were such
that the magnitude $\left\vert \delta q_{\mathbf{k}r}(t_{f})\right\vert $ of
the final displacement were small compared to $\Delta q_{\mathbf{k}r}(t_{f})$,
then relaxation (with respect to $q_{\mathbf{k}r}$) during the interval
$[t_{i},t_{f}]$ would in general be impossible, as the configurations would
not move far enough for the two `fluids' $P$ and $|\Psi|^{2}$ to be
significantly `stirred' or mixed (with respect to $q_{\mathbf{k}r}$). This is
clear because the time evolutions of $P$ and $|\Psi|^{2}$ are determined by
the same continuity equation and the same family of trajectories. For example,
if $|\Psi|^{2}$ is initially spread over an interval $[a,b]$ of $q_{\mathbf{k}%
r}$-space of length $\sim\Delta q_{\mathbf{k}r}(t_{i})$, and if the
displacements of all the trajectories during $[t_{i},t_{f}]$ are indeed such
that $\left\vert \delta q_{\mathbf{k}r}(t_{f})\right\vert <<\Delta
q_{\mathbf{k}r}(t_{f})$, then $|\Psi|^{2}$ will essentially remain spread over
$[a,b]$ during $[t_{i},t_{f}]$ (with $\Delta q_{\mathbf{k}r}(t_{f}%
)\approx\Delta q_{\mathbf{k}r}(t_{i})$); while if $P$ is, say, initially
confined to the left half of the interval $[a,b]$, it will essentially remain
so during $[t_{i},t_{f}]$, and there will be no significant evolution towards
equilibrium (for the coordinate $q_{\mathbf{k}r}$).

Thus we might take our condition to be $\left\vert \delta q_{\mathbf{k}%
r}(t_{f})\right\vert <<\Delta q_{\mathbf{k}r}(t_{f})$. However, if there were
some isolated trajectories for which $\left\vert \delta q_{\mathbf{k}r}%
(t_{f})\right\vert \sim\Delta q_{\mathbf{k}r}(t_{f})$, or even $\left\vert
\delta q_{\mathbf{k}r}(t_{f})\right\vert \gtrsim\Delta q_{\mathbf{k}r}(t_{f}%
)$, while most trajectories still satisfied $\left\vert \delta q_{\mathbf{k}%
r}(t_{f})\right\vert <<\Delta q_{\mathbf{k}r}(t_{f})$ (where `most' could be
defined for example with respect to the Lebesgue measure or with respect to
the $|\Psi|^{2}$-measure), then relaxation would still be impossible in
general. Hence we may take the weaker condition%
\begin{equation}
\left\langle \left\vert \delta q_{\mathbf{k}r}(t_{f})\right\vert \right\rangle
_{\mathrm{eq}}<<\Delta q_{\mathbf{k}r}(t_{f})\ , \label{condn0}%
\end{equation}
where $\left\langle \left\vert \delta q_{\mathbf{k}r}(t_{f})\right\vert
\right\rangle _{\mathrm{eq}}$ is the average of $\left\vert \delta
q_{\mathbf{k}r}(t_{f})\right\vert $ over an equilibrium ensemble.

The condition (\ref{condn0}) implies that `most' of the ensemble cannot move
by `much' more than a small fraction of $\Delta q_{\mathbf{k}r}(t_{f})$, in
the following precise sense. Define $\delta\equiv\left\vert \delta
q_{\mathbf{k}r}(t_{f})\right\vert =\left\vert q_{\mathbf{k}r}(t_{f}%
)-q_{\mathbf{k}r}(t_{i})\right\vert \geq0$ (where $\delta=\delta(q_{i},t_{f})$
is a function of the initial total configuration $q_{i}\equiv q(t_{i})$). From
(\ref{condn0}), we can write $\left\langle \delta\right\rangle _{\mathrm{eq}%
}<\varepsilon\Delta q_{\mathbf{k}r}(t_{f})$ for some $\varepsilon<<1$. We can
then show that `most' values of $\delta$ cannot be `much' bigger than
$\varepsilon\Delta q_{\mathbf{k}r}(t_{f})$ --- where we define `most' with
respect to the equilibrium measure $|\Psi\lbrack q_{i},t_{i}]|^{2}dq_{i}$ over
the ensemble of initial configurations $q_{i}$, and where we define $\delta$
to be `much' bigger than $\varepsilon\Delta q_{\mathbf{k}r}(t_{f})$ if
$\delta>2\varepsilon\Delta q_{\mathbf{k}r}(t_{f})$. Let $R$ be the set of
initial points $q_{i}$ such that $\delta>\varepsilon\Delta q_{\mathbf{k}%
r}(t_{f})+d$, for some fixed $d>0$. Such points make up a certain fraction $F$
of the ensemble, that is $F=\int_{R}dq_{i}\ |\Psi\lbrack q_{i},t_{i}]|^{2}$.
We have a mean%
\[
\left\langle \delta\right\rangle _{\mathrm{eq}}=\int dq_{i}\ |\Psi\lbrack
q_{i},t_{i}]|^{2}.\delta(q_{i},t_{f})\ .
\]
Since $\delta\geq0$ for all $q_{i}$, we have%
\begin{align*}
\left\langle \delta\right\rangle _{\mathrm{eq}}  &  \geq\int_{R}dq_{i}%
\ |\Psi\lbrack q_{i},t_{i}]|^{2}.\delta(q_{i},t_{f})\\
&  >\int_{R}dq_{i}\ |\Psi\lbrack q_{i},t_{i}]|^{2}.\left(  \varepsilon\Delta
q_{\mathbf{k}r}(t_{f})+d\right)  =F.\left(  \varepsilon\Delta q_{\mathbf{k}%
r}(t_{f})+d\right)  \ .
\end{align*}
Given (\ref{condn0}), or $\left\langle \delta\right\rangle _{\mathrm{eq}%
}<\varepsilon\Delta q_{\mathbf{k}r}(t_{f})$, we then have $\varepsilon\Delta
q_{\mathbf{k}r}(t_{f})>F.\left(  \varepsilon\Delta q_{\mathbf{k}r}%
(t_{f})+d\right)  $ and so%
\begin{equation}
d<\frac{(1-F)}{F}\varepsilon\Delta q_{\mathbf{k}r}(t_{f})\ . \label{delta}%
\end{equation}
Now, if $F>\frac{1}{2}$ (that is, if `most' initial points yield
$\delta>\varepsilon\Delta q_{\mathbf{k}r}(t_{f})+d$), then $(1-F)/F<1$ and so
$d<\varepsilon\Delta q_{\mathbf{k}r}(t_{f})$. We may then indeed conclude that
`most' of the initial ensemble cannot move by `much' more than $\varepsilon
\Delta q_{\mathbf{k}r}(t_{f})$. In this case, even an approximate relaxation
cannot (in general) occur.

If (\ref{condn0}) is satisfied, then, relaxation will in general be
suppressed. Of course, while (\ref{condn0}) is a sufficient condition for
relaxation suppression, it is not necessary: in principle, the trajectories
could even wander over distances larger than $\Delta q_{\mathbf{k}r}(t_{f})$
but without a sufficiently complex flow to drive the ensemble towards
equilibrium. (As discussed in section 7, it is reasonable to assume that this
is unlikely.)

While (\ref{condn0}) provides a condition for the freezing of quantum
nonequilibrium, in practice it is likely to be more stringent than is
necessary. Without attempting to give a rigorous justification, we expect that
there will be cases where the weaker condition%
\begin{equation}
\left\langle \left\vert \delta q_{\mathbf{k}r}(t_{f})\right\vert \right\rangle
_{\mathrm{eq}}<\Delta q_{\mathbf{k}r}(t_{f}) \label{condn01}%
\end{equation}
suffices to prevent relaxation, at least partially (that is, some significant
relaxation towards equilibrium will occur but significant deviations from
equilibrium will remain). Generally speaking, we expect that the transition
from essentially complete relaxation suppression to essentially full
relaxation will take place when the ratio $r\equiv\left\langle \left\vert
\delta q_{\mathbf{k}r}(t_{f})\right\vert \right\rangle _{\mathrm{eq}}/\Delta
q_{\mathbf{k}r}(t_{f})$ increases from $r<<1$ to $r\gtrsim1$, with the
critical demarcation line being somewhere in the neigbourhood of $r\sim1$. We
therefore expect that the weaker condition (\ref{condn01}) will define
(approximately) essentially the whole of the suppression regime, including
those cases where significant relaxation towards equilibrium does occur but
where significant deviations from equilibrium still remain. (Note that
(\ref{condn01}) implies that `most' of the ensemble cannot move by `much' more
than $\Delta q_{\mathbf{k}r}(t_{f})$, in the sense given above.)

Pending a more precise treatment, then, here we shall take (\ref{condn01}) as
our condition for the freezing --- or at least partial freezing --- of quantum nonequilibrium.

Let us now proceed to draw inferences from (\ref{condn01}). Note first that
the final displacement $\delta q_{\mathbf{k}r}(t_{f})$ has modulus $\left\vert
\delta q_{\mathbf{k}r}(t_{f})\right\vert \leq\int_{t_{i}}^{t_{f}%
}dt\ \left\vert \dot{q}_{\mathbf{k}r}(t)\right\vert $ (where $\int_{t_{i}%
}^{t_{f}}dt\ \left\vert \dot{q}_{\mathbf{k}r}(t)\right\vert $ is the path
length). The equilibrium mean $\left\langle \left\vert \delta q_{\mathbf{k}%
r}(t_{f})\right\vert \right\rangle _{\mathrm{eq}}$ then satisfies%
\begin{equation}
\left\langle \left\vert \delta q_{\mathbf{k}r}(t_{f})\right\vert \right\rangle
_{\mathrm{eq}}\leq\left\langle \int_{t_{i}}^{t_{f}}dt\ \left\vert \dot
{q}_{\mathbf{k}r}(t)\right\vert \right\rangle _{\mathrm{eq}}=\int_{t_{i}%
}^{t_{f}}dt\ \left\langle \left\vert \dot{q}_{\mathbf{k}r}(t)\right\vert
\right\rangle _{\mathrm{eq}}\ , \label{intsp}%
\end{equation}
where the equilibrium mean speed $\left\langle \left\vert \dot{q}%
_{\mathbf{k}r}(t)\right\vert \right\rangle _{\mathrm{eq}}$ at time $t$ is%
\begin{equation}
\left\langle \left\vert \dot{q}_{\mathbf{k}r}(t)\right\vert \right\rangle
_{\mathrm{eq}}=\int dq\ |\Psi\lbrack q,t]|^{2}|\dot{q}_{\mathbf{k}r}(q,t)|
\end{equation}
(the velocity $\dot{q}_{\mathbf{k}r}(q,t)$ being given by (\ref{deB2}) as a
time-dependent function of the total configuration $q$).

For the sake of clarity, let us explicitly demonstrate the last equality in
(\ref{intsp}). The initial equilibrium distribution $|\Psi\lbrack q_{i}%
,t_{i}]|^{2}$ represents an ensemble of initial (total) configurations $q_{i}%
$. From each $q_{i}$, the de Broglie velocity field generates a trajectory
$q(t)$ (for the whole system), and each such trajectory implies a subsystem
trajectory $q_{\mathbf{k}r}(t)$. Thus, at any time $t$, the subsystem velocity
$\dot{q}_{\mathbf{k}r}$ may be regarded as a function of $q_{i}$ and of $t$
(assuming the wave functional is given). We may then write $\dot
{q}_{\mathbf{k}r}=\dot{q}_{\mathbf{k}r}(q_{i},t)$ --- where of course $\dot
{q}_{\mathbf{k}r}(q_{i},t)$ and $\dot{q}_{\mathbf{k}r}(q,t)$ here denote two
different functions of the first argument. (This notation is strictly speaking
ambiguous, but clear from the context.) We then have%
\begin{equation}
\left\langle \left\vert \dot{q}_{\mathbf{k}r}(t)\right\vert \right\rangle
_{\mathrm{eq}}=\int dq_{i}\ |\Psi\lbrack q_{i},t_{i}]|^{2}|\dot{q}%
_{\mathbf{k}r}(q_{i},t)|
\end{equation}
(with the mean taken over the distribution $|\Psi\lbrack q_{i},t_{i}]|^{2}$ at
the fixed \textit{initial} time $t_{i}$), so that%
\begin{align*}
\int_{t_{i}}^{t_{f}}dt\ \left\langle \left\vert \dot{q}_{\mathbf{k}%
r}(t)\right\vert \right\rangle _{\mathrm{eq}}  &  =\int dq_{i}\ |\Psi\lbrack
q_{i},t_{i}]|^{2}\left(  \int_{t_{i}}^{t_{f}}dt\ |\dot{q}_{\mathbf{k}r}%
(q_{i},t)|\right) \\
&  =\left\langle \int_{t_{i}}^{t_{f}}dt\ \left\vert \dot{q}_{\mathbf{k}%
r}(t)\right\vert \right\rangle _{\mathrm{eq}}\ ,
\end{align*}
as used above. (We have shifted notation back and forth, with $\dot
{q}_{\mathbf{k}r}(t)$ and $\dot{q}_{\mathbf{k}r}(q_{i},t)$ denoting the same thing.)

Using $\left\langle x\right\rangle \leq\sqrt{\left\langle x^{2}\right\rangle
}$ for any $x$, we then have%
\[
\left\langle \left\vert \delta q_{\mathbf{k}r}(t_{f})\right\vert \right\rangle
_{\mathrm{eq}}\leq\int_{t_{i}}^{t_{f}}dt\ \sqrt{\left\langle \left\vert
\dot{q}_{\mathbf{k}r}(t)\right\vert ^{2}\right\rangle _{\mathrm{eq}}}\ .
\]

Now note that, at any time $t$,%
\begin{align}
a^{6}\left\langle |\dot{q}_{\mathbf{k}r}|^{2}\right\rangle _{\mathrm{eq}}  &
=\left\langle \left(  \frac{\partial S}{\partial q_{\mathbf{k}r}}\right)
^{2}\right\rangle _{\mathrm{eq}}=\int dq\ |\Psi\lbrack q,t]|^{2}\left(
\frac{\partial S[q,t]}{\partial q_{\mathbf{k}r}}\right)  ^{2}\nonumber\\
&  =\left\langle \hat{\pi}_{\mathbf{k}r}^{2}\right\rangle -\int dq\ \left(
\frac{\partial|\Psi\lbrack q,t]|}{\partial q_{\mathbf{k}r}}\right)  ^{2}
\label{eq1}%
\end{align}
(where $\left\langle \hat{\Omega}\right\rangle $ denotes the usual quantum
expectation value for an operator $\hat{\Omega}$). The last equality follows
from%
\[
\left\langle \hat{\pi}_{\mathbf{k}r}^{2}\right\rangle =\int dq\ \Psi^{\ast
}\left(  -\frac{\partial^{2}}{\partial q_{\mathbf{k}r}^{2}}\right)  \Psi=\int
dq\ \frac{\partial\Psi^{\ast}}{\partial q_{\mathbf{k}r}}\frac{\partial\Psi
}{\partial q_{\mathbf{k}r}}\ ,
\]
and from%
\[
\frac{\partial\Psi^{\ast}}{\partial q_{\mathbf{k}r}}\frac{\partial\Psi
}{\partial q_{\mathbf{k}r}}=\left(  \frac{\partial|\Psi|}{\partial
q_{\mathbf{k}r}}\right)  ^{2}+|\Psi|^{2}\left(  \frac{\partial S}{\partial
q_{\mathbf{k}r}}\right)  ^{2}\ .
\]
Thus, since $(\partial|\Psi|/\partial q_{\mathbf{k}r})^{2}\geq0$, we have%
\begin{equation}
a^{6}\left\langle |\dot{q}_{\mathbf{k}r}|^{2}\right\rangle _{\mathrm{eq}}%
\leq\left\langle \hat{\pi}_{\mathbf{k}r}^{2}\right\rangle \ , \label{ineqAV}%
\end{equation}
and so%
\begin{equation}
\left\langle \left\vert \delta q_{\mathbf{k}r}(t_{f})\right\vert \right\rangle
_{\mathrm{eq}}\leq\int_{t_{i}}^{t_{f}}dt\ \frac{1}{a^{3}}\sqrt{\left\langle
\hat{\pi}_{\mathbf{k}r}^{2}\right\rangle }%
\end{equation}
(where it is understood that quantities under the integral sign are evaluated
at time $t$).

Since $\left\langle \hat{q}_{\mathbf{k}r}^{2}\right\rangle >0$, we also have%
\begin{equation}
\left\langle \hat{\pi}_{\mathbf{k}r}^{2}\right\rangle <2a^{3}\left\langle
\hat{H}_{\mathbf{k}r}\right\rangle \ , \label{ineq2}%
\end{equation}
and so%
\[
\left\langle \left\vert \delta q_{\mathbf{k}r}(t_{f})\right\vert \right\rangle
_{\mathrm{eq}}<\int_{t_{i}}^{t_{f}}dt\ \frac{1}{a^{3}}\sqrt{2a^{3}\left\langle
\hat{H}_{\mathbf{k}r}\right\rangle }\ .
\]

Introducing the number operator $\hat{n}_{\mathbf{k}r}$, where $\left\langle
\hat{n}_{\mathbf{k}r}\right\rangle \geq0$, the mean energy in the mode
$\mathbf{k}r$ is%
\[
\left\langle \hat{H}_{\mathbf{k}r}\right\rangle =(\left\langle \hat
{n}_{\mathbf{k}r}\right\rangle +\frac{1}{2})\frac{k}{a}\ .
\]
We then have%
\begin{equation}
\left\langle \left\vert \delta q_{\mathbf{k}r}(t_{f})\right\vert \right\rangle
_{\mathrm{eq}}<\int_{t_{i}}^{t_{f}}dt\ \frac{1}{a^{2}}\sqrt{2k(\left\langle
\hat{n}_{\mathbf{k}r}\right\rangle +1/2)}\ . \label{+1}%
\end{equation}

The mean $\left\langle \left\vert \delta q_{\mathbf{k}r}(t_{f})\right\vert
\right\rangle _{\mathrm{eq}}$ at time $t_{f}$ is to be compared with the width
$\Delta q_{\mathbf{k}r}(t_{f})$ (with respect to $q_{\mathbf{k}r}$) of the
quantum distribution $|\Psi\lbrack q,t_{f}]|^{2}$ at time $t_{f}$. Using the
uncertainty relation $\Delta q_{\mathbf{k}r}\Delta\pi_{\mathbf{k}r}\geq
\frac{1}{2}$ and $\Delta\pi_{\mathbf{k}r}\leq\sqrt{\left\langle \hat{\pi
}_{\mathbf{k}r}^{2}\right\rangle }$, we have $1/\Delta q_{\mathbf{k}r}%
\leq2\sqrt{\left\langle \hat{\pi}_{\mathbf{k}r}^{2}\right\rangle }$. Again
using (\ref{ineq2}) we then have%
\begin{equation}
1/\Delta q_{\mathbf{k}r}<2\sqrt{2a^{3}\left\langle \hat{H}_{\mathbf{k}%
r}\right\rangle }=2a\sqrt{2k(\left\langle \hat{n}_{\mathbf{k}r}\right\rangle
+1/2)}\ . \label{+2}%
\end{equation}

Combining the results (\ref{+1}) and (\ref{+2}), we obtain an upper bound for
the ratio%
\begin{equation}
\frac{\left\langle \left\vert \delta q_{\mathbf{k}r}(t_{f})\right\vert
\right\rangle _{\mathrm{eq}}}{\Delta q_{\mathbf{k}r}(t_{f})}<4ka_{f}%
\sqrt{\left\langle \hat{n}_{\mathbf{k}r}\right\rangle _{f}+1/2}\int_{t_{i}%
}^{t_{f}}dt\ \frac{1}{a^{2}}\sqrt{\left\langle \hat{n}_{\mathbf{k}%
r}\right\rangle +1/2} \label{+}%
\end{equation}
(where $a_{f}\equiv a(t_{f})$, and so on). Note that $\left\langle \hat
{n}_{\mathbf{k}r}\right\rangle $ is in general a function of time $t$, and
that the inequality (\ref{+}) holds for any arbitrary (in general entangled)
state $\Psi$.

We may now consider the following inequality, that the right-hand side of
(\ref{+}) is less than one, that is%
\begin{equation}
\frac{1}{k}>4a_{f}\sqrt{\left\langle \hat{n}_{\mathbf{k}r}\right\rangle
_{f}+1/2}\int_{t_{i}}^{t_{f}}dt\ \frac{1}{a^{2}}\sqrt{\left\langle \hat
{n}_{\mathbf{k}r}\right\rangle +1/2}\ . \label{condn}%
\end{equation}
When this `freezing inequality' is satisfied, $\left\langle \left\vert \delta
q_{\mathbf{k}r}(t_{f})\right\vert \right\rangle _{\mathrm{eq}}/\Delta
q_{\mathbf{k}r}(t_{f})<1$ and initial quantum nonequilibrium will be (at least
partially) `frozen'.

We may also write (\ref{condn}) directly in terms of $\left\langle \hat
{H}_{\mathbf{k}r}\right\rangle $, yielding%
\begin{equation}
4a_{f}\sqrt{a_{f}\left\langle \hat{H}_{\mathbf{k}r}\right\rangle _{f}}%
\int_{t_{i}}^{t_{f}}dt\ \frac{1}{a^{2}}\sqrt{a\left\langle \hat{H}%
_{\mathbf{k}r}\right\rangle }<1\ . \label{condn2}%
\end{equation}
The dependence on the wave number $k$ is of course still present in
$\left\langle \hat{H}_{\mathbf{k}r}\right\rangle $. Roughly speaking, the
freezing inequality (\ref{condn2}) requires that the mean energy $\left\langle
\hat{H}_{\mathbf{k}r}\right\rangle $ in the mode $\mathbf{k}r$ be not too
large over the time interval $[t_{i},t_{f}]$ (see below).

\section{Generalisations}

Before discussing the consequences of the above results, let us first
generalise them to more realistic situations. The above derivation of the
freezing inequality (\ref{condn}) (or (\ref{condn2})) assumed that the quantum
state was pure and that the field was free. The derivation is easily
generalised to mixed states and to (finite models of) interacting fields.

With these generalisations in hand, one can then discuss nonequilibrium
freezing for a mixed (for example thermal) ensemble of interacting particles,
and one can apply the results to realistic models of the early universe.

\subsection{Mixed States}

In quantum theory, a mixed state is represented by a density operator
$\hat{\rho}$, which may be written as a decomposition%
\begin{equation}
\hat{\rho}=\sum_{\alpha}p_{\alpha}|\Psi_{\alpha}\rangle\langle\Psi_{\alpha
}|\ , \label{rhodec}%
\end{equation}
with appropriate probability weights $p_{\alpha}$ and pure states
$|\Psi_{\alpha}\rangle$. For a scalar field $\phi$, the quantum-theoretical
distribution for $\phi$ will be%
\begin{equation}
P_{\mathrm{QT}}[\phi,t]=\langle\phi|\hat{\rho}(t)|\phi\rangle=\sum_{\alpha
}p_{\alpha}|\Psi_{\alpha}[\phi,t]|^{2}\ . \label{Peq}%
\end{equation}
The decomposition of $\hat{\rho}$ is generally non-unique, and different
decompositions of the same $\hat{\rho}$ are physically equivalent in all respects.

The situation is different in pilot-wave theory. A mixed quantum state is
interpreted as a statistical mixture of physically-real pilot waves
$\Psi_{\alpha}$, with probability weights $p_{\alpha}$, corresponding to a
preferred decomposition of $\hat{\rho}$ \cite{BH96}. For a given element of
the ensemble, the de Broglian velocity of the actual configuration is
determined by the actual pilot wave $\Psi_{\alpha}$. A different decomposition
of $\hat{\rho}$ would generally yield different velocities, and so be
physically distinct at the fundamental level. (Note that, in quantum
nonequilibrium, the velocities and trajectories for single systems can be
measured without necessarily disturbing the wave functions \cite{AVPr02,PV06},
enabling the preferred decomposition to be detected. The operational
equivalence of different decompositions of $\hat{\rho}$ is a peculiarity of
the quantum equilibrium state; see ref. \cite{AVBHs}.)

Now, given such a preferred decomposition, for each pure subensemble with wave
functional $\Psi_{\alpha}[\phi,t]$ --- taking the system to consist of a
scalar field $\phi$ --- we may define a distribution $P_{\alpha}[\phi,t]$
(generally $\neq|\Psi_{\alpha}[\phi,t]|^{2}$) and an associated $H$-function%
\[
H_{\alpha}=\int D\phi\ P_{\alpha}\ln(P_{\alpha}/|\Psi_{\alpha}|^{2})
\]
(for some appropriate measure $D\phi$). The whole ensemble has a distribution%
\begin{equation}
P[\phi,t]=\sum_{\alpha}p_{\alpha}P_{\alpha}[\phi,t]\ , \label{Pnoneq}%
\end{equation}
and the mean $H$-function%
\begin{equation}
H=\sum_{\alpha}p_{\alpha}H_{\alpha}%
\end{equation}
obeys a coarse-graining $H$-theorem (for a closed system with constant
$p_{\alpha}$) \cite{AVBHs}. The equilibrium minimum $H=0$ (which may be
approached in a coarse-grained sense) corresponds to $H_{\alpha}=0$ and
$P_{\alpha}=|\Psi_{\alpha}|^{2}$ for every $\alpha$, so that (\ref{Pnoneq})
reduces to (\ref{Peq}).

Thus, we may discuss relaxation for a mixed state in terms of relaxation for
its component pure subensembles. We may then consider the freezing inequality
(\ref{condn}) (or (\ref{condn2})) for each pure subensemble separately.
Clearly, the inequality might hold for some subensembles and not for others
(or for all of them, or none).

If the (quantum) mean occupation number for state $|\Psi_{\alpha}\rangle$ is
$\left\langle \hat{n}_{\mathbf{k}r}\right\rangle _{\alpha}\equiv\left\langle
\Psi_{\alpha}\right\vert \hat{n}_{\mathbf{k}r}\left\vert \Psi_{\alpha
}\right\rangle $, then for a mixed state (\ref{rhodec}) the overall mean
occupation number will be%
\[
\overline{\left\langle \hat{n}_{\mathbf{k}r}\right\rangle }=\sum_{\alpha
}p_{\alpha}\left\langle \hat{n}_{\mathbf{k}r}\right\rangle _{\alpha}\ .
\]
For example, for a thermal ensemble with temperature $T$, we will have the
Planck distribution%
\[
\overline{\left\langle \hat{n}_{\mathbf{k}r}\right\rangle }_{\mathrm{P}}%
=\frac{1}{e^{\hbar\omega/k_{\mathrm{B}}T}-1}\ .
\]

In general, $\left\langle \hat{n}_{\mathbf{k}r}\right\rangle _{\alpha}$ for a
pure subensemble will differ from $\overline{\left\langle \hat{n}%
_{\mathbf{k}r}\right\rangle }$, and the total ensemble will contain a range of
different values for $\left\langle \hat{n}_{\mathbf{k}r}\right\rangle
_{\alpha}$. Initial quantum nonequilibrium will be frozen for the pure
subensemble with wave functional $\Psi_{\alpha}$, if the corresponding
quantity $\left\langle \hat{n}_{\mathbf{k}r}\right\rangle _{\alpha}$ satisfies
the freezing inequality (\ref{condn}) (with $\left\langle \hat{n}%
_{\mathbf{k}r}\right\rangle $ replaced by $\left\langle \hat{n}_{\mathbf{k}%
r}\right\rangle _{\alpha}$).

To investigate which (if any) pure subensembles will satisfy the freezing
inequality (\ref{condn}), we need to know the quantities $\left\langle \hat
{n}_{\mathbf{k}r}\right\rangle _{\alpha}$ as functions of time, that is, we
need to know which pure states $\left\vert \Psi_{\alpha}\right\rangle $ are
present in the total ensemble. Despite the operational equivalence of
different decompositions of $\hat{\rho}$ in quantum theory, it has been argued
that, in the case of thermal (canonical) ensembles, there is a natural
probability measure on the space of normalised wave functions, the `Gaussian
adjusted projected measure', which is unique for each $\hat{\rho}$, and which
may be used to define a preferred decomposition \cite{Gold}. This proposal has
been applied to the case of an ideal gas (though described in terms of
particle theory rather than field theory) \cite{Tum}. For our purposes, we
would need to apply the preferred measure to a thermal ensemble of wave
functionals in field theory on expanding space, and use the results to deduce
which (if any) subensembles of finite measure satisfy the freezing inequality
(\ref{condn}) (or (\ref{condn2})). We do not attempt such a calculation here,
but it should be clear that the problem is well-defined.

If certain pure subensembles --- with labels $\alpha$ in some set $S$ --- are
predicted to be frozen, then (assuming initial nonequilibrium) the total
ensemble distribution of $\phi$ will take the form%
\[
P[\phi,t]=\sum_{\alpha\in S}p_{\alpha}P_{\alpha}[\phi,t]+\sum_{\alpha\notin
S}p_{\alpha}|\Psi_{\alpha}[\phi,t]|^{2}\ ,
\]
where $P_{\alpha}\neq|\Psi_{\alpha}|^{2}$ (for $\alpha\in S$), and $P[\phi,t]$
will generally differ from the equilibrium result (\ref{Peq}).

The physics of nonequilibrium mixed states needs further development. In
particular, one should explore how measurements could probe the nonequilibrium
physics particular to a specific pure subensemble (noting again that, unlike
in quantum theory, in nonequilibrium pilot-wave theory it is operationally
meaningful to speak of the physics of component pure subensembles). However,
the above suffices for the purposes of this paper.

\subsection{Interacting Fields}

Our derivation in section 4 of the freezing inequality (\ref{condn}) assumed
that the field $\phi$ was free. The derivation is easily generalised to
interacting fields, at least if one considers finite models with an
appropriate high-frequency cutoff (so that divergences may be ignored).

Let the scalar field $\phi$ interact with other fields, denoted collectively
by $\Phi$. (These other fields need not be scalars.) We have a total
Hamiltonian%
\[
\hat{H}_{\mathrm{total}}=\hat{H}+\hat{H}_{\Phi}+\hat{H}_{\mathrm{I}}\ ,
\]
where $\hat{H}$ and $\hat{H}_{\Phi}$ are respectively the free Hamiltonians
for $\phi$ and $\Phi$, while $\hat{H}_{\mathrm{I}}$ is the interaction Hamiltonian.

We may still of course write $\phi$ in terms of its Fourier components
$\phi_{\mathbf{k}r}$, and the free Hamiltonian $\hat{H}$ still decomposes into
a sum $\hat{H}=\sum_{\mathbf{k}r}\hat{H}_{\mathbf{k}r}$, with $\hat
{H}_{\mathbf{k}r}=(\hat{n}_{\mathbf{k}r}+\frac{1}{2})\frac{k}{a}$, exactly as
before. Equation (\ref{eq1}) still holds (for a pure subensemble with wave
functional $\Psi$, and where the total configuration $q$ now includes $\Phi$
as well as $\phi$). So we still have the inequality (\ref{ineqAV}). The other
inequalities --- such as (\ref{ineq2}) and $\Delta q_{\mathbf{k}r}\Delta
\pi_{\mathbf{k}r}\geq\frac{1}{2}$ --- are also valid as in the case of a free
field. We therefore arrive again at the upper bound (\ref{+}) and the freezing
inequalities (\ref{condn}) and (\ref{condn2}).

The only difference from the free case is in the time evolution of
$\left\langle \hat{n}_{\mathbf{k}r}\right\rangle $ (or of $\left\langle
\hat{H}_{\mathbf{k}r}\right\rangle $), which now involves contributions from
$\hat{H}_{I}$:%
\[
\frac{d\left\langle \hat{n}_{\mathbf{k}r}\right\rangle }{dt}=\left\langle
\frac{\partial\hat{n}_{\mathbf{k}r}}{\partial t}\right\rangle -i\left\langle
[\hat{n}_{\mathbf{k}r},\hat{H}_{I}]\right\rangle \ .
\]
The calculation of $\left\langle \hat{n}_{\mathbf{k}r}\right\rangle _{t}$ as a
function of time $t$ will then be more complicated than in the free case,
where only the first term appears on the right hand side. (The evolution of
$\left\langle \hat{n}_{\mathbf{k}r}\right\rangle _{t}$ in the free case is
studied in section 8.)

\section{General Implications of the Freezing Inequality}

Quite generally, then, even for an interacting field in a mixed state, we may
conclude that relaxation will be suppressed --- that is, nonequilibrium will
be frozen --- for modes whose (time-dependent) mean occupation number
$\left\langle \hat{n}_{\mathbf{k}r}\right\rangle $ satisfies the inequality
(\ref{condn}).

For a given time evolution, defined by $a(t)$ and $\left\langle \hat
{n}_{\mathbf{k}r}\right\rangle _{t}$ (for all $\mathbf{k}r$) on $[t_{i}%
,t_{f}]$, it is of course possible that (\ref{condn}) will not be satisfied
for any value of $k$, and that all modes relax (at least approximately)
towards equilibrium during the interval $[t_{i},t_{f}]$. On the other hand if,
for a given time evolution, (\ref{condn}) is satisfied only for certain values
of $k$, then we can predict that significant deviations from quantum
equilibrium are to be expected only for those particular values of $k$.

We emphasise that, for each mode, whether or not the inequality (\ref{condn})
is satisfied depends on the history of the expansion and on the time evolution
of the quantum state of the field.

For a radiation-dominated expansion on $[t_{i},t_{f}]$, with $a(t)=a_{f}%
(t/t_{f})^{1/2}$, we may make a general statement about the kind of modes that
can satisfy (\ref{condn}): the physical wavelength $\lambda_{\mathrm{phys}%
}(t_{f})=a_{f}(2\pi/k)$ at time $t_{f}$ must be larger than the Hubble radius
$H_{f}^{-1}$ at time $t_{f}$ (assuming that $t_{f}\gtrsim(1.17)t_{i}$).

This is easily shown for any quantum state. Since $\left\langle \hat
{n}_{\mathbf{k}r}\right\rangle \geq0$, the inequality (\ref{condn}) (assuming
it to hold) implies that%
\[
\frac{1}{k}>2a_{f}\int_{t_{i}}^{t_{f}}dt\ \frac{1}{a^{2}}=\frac{2t_{f}}{a_{f}%
}\ln(t_{f}/t_{i})\ ,
\]
or%
\begin{equation}
\lambda_{\mathrm{phys}}(t_{f})>2\pi H_{f}^{-1}\ln(t_{f}/t_{i})\ ,
\label{lam-con}%
\end{equation}
where $H_{f}^{-1}=2t_{f}$ and where the right-hand side is indeed larger than
$H_{f}^{-1}$ if $t_{f}>t_{i}\exp(1/2\pi)\simeq(1.17)t_{i}$. (This is of course
not to suggest that the freezing inequality is satisfied for all super-Hubble
modes: rather, if the inequality is satisfied, then the corresponding modes
must be super-Hubble.)

Note that, in any reasonable application of this result, the factor $\ln
(t_{f}/t_{i})$ will not be large. For example, taking $t_{i}\sim
t_{\mathrm{P}}\sim10^{-43}\ \mathrm{s}$, for $t_{f}\sim10^{-35}\ \mathrm{s}$
(the time at which inflation begins in some models \cite{Pad93}) we have
$\ln(t_{f}/t_{i})\sim\ln10^{8}\sim20$, while even for $t_{f}\sim1\ \mathrm{s}$
(the time of neutrino decoupling) we have $\ln(t_{f}/t_{i})\sim\ln10^{43}%
\sim10^{2}$. The factor $2\pi\ln(t_{f}/t_{i})$ is then likely to be at most of
order $10^{2}-10^{3}$, in which case the minimal value of $\lambda
_{\mathrm{phys}}(t_{f})$ for nonequilibrium field modes will be at most two or
three orders of magnitude larger than the Hubble radius $H_{f}^{-1}$.

On the other hand, again for a radiation-dominated expansion, the true lower
bound on $\lambda_{\mathrm{phys}}(t_{f})$ (set by (\ref{condn})) will be much
larger than $2\pi H_{f}^{-1}\ln(t_{f}/t_{i})$ if $\left\langle \hat
{n}_{\mathbf{k}r}\right\rangle _{t}>>1$ during the period $[t_{i},t_{f}]$, as
is clear from (\ref{condn}).

Thus, de Broglie-Bohm theory (with the assumption of early quantum
nonequilibrium at some initial time $t_{i}$) predicts that residual or
`frozen' nonequilibrium will exist at later times $t_{f}>t_{i}$ for modes
satisfying the inequality (\ref{condn}), where for a radiation-dominated
expansion the physical wavelength $\lambda_{\mathrm{phys}}(t_{f})$ of
nonequilibrium modes at time $t_{f}$ must be bigger than $2\pi H_{f}^{-1}%
\ln(t_{f}/t_{i})$ (at least).

If we take the freezing inequality in the form (\ref{condn2}), we see that,
roughly speaking, it entails an upper bound on the mean energy $\left\langle
\hat{H}_{\mathbf{k}r}\right\rangle $ per mode over time. More precisely if,
for example, $\left\langle \hat{H}_{\mathbf{k}r}\right\rangle \geq\left\langle
\hat{H}_{\mathbf{k}r}\right\rangle _{\min}$ throughout $[t_{i},t_{f}]$, then
(\ref{condn2}) implies that%
\begin{equation}
\left\langle \hat{H}_{\mathbf{k}r}\right\rangle _{\min}<\frac{1}{4a_{f}%
^{3/2}\int_{t_{i}}^{t_{f}}dt\ a^{-3/2}}\ .
\end{equation}
For a radiation-dominated expansion, and assuming $t_{f}/t_{i}>>1$, we then
have (inserting $\hbar$)%
\begin{equation}
\left\langle \hat{H}_{\mathbf{k}r}\right\rangle _{\min}<\frac{\hbar}{16t_{f}%
}=\frac{\hbar}{8H_{f}^{-1}} \label{hbar}%
\end{equation}
(where, dimensionally speaking, $H_{f}^{-1}=2t_{f}$ is the Hubble time and
$cH_{f}^{-1}$ is the Hubble radius).

Finally, we note that \textit{violation} of the freezing inequality
(\ref{condn}) in the infra-red limit $k\rightarrow0$ requires that
$\left\langle \hat{n}_{\mathbf{k}r}\right\rangle $ be divergent as
$k\rightarrow0$. Alternatively, for (\ref{condn2}) to be violated as
$k\rightarrow0$, the mean energy per mode $\left\langle \hat{H}_{\mathbf{k}%
r}\right\rangle $ must remain finite as $k\rightarrow0$.

\section{Relaxation for Modes Violating the Freezing Inequality}

We have shown that, for modes satisfying (\ref{condn}), relaxation will be
suppressed over the time interval $[t_{i},t_{f}]$. For a radiation-dominated
expansion we know from (\ref{lam-con}) that such modes, if they exist, must
have super-Hubble wavelengths. Further, as discussed in section 3, we know
from previous studies that relaxation is likely to occur in the
short-wavelength (Minkowski) limit. What can we say about modes that violate
the freezing inequality (\ref{condn}) without approaching the Minkowski limit?

Our derivation of the upper bound (\ref{+}) made use of several general
inequalities (such as $\left\langle \hat{\pi}_{\mathbf{k}r}^{2}\right\rangle
\leq2a^{3}\left\langle \hat{H}_{\mathbf{k}r}\right\rangle $). For a large
class of quantum states, these general inequalities could be replaced by
approximate equalities, to be used as rough, order-of-magnitude estimates (for
example, $\left\langle \hat{\pi}_{\mathbf{k}r}^{2}\right\rangle \sim
2a^{3}\left\langle \hat{H}_{\mathbf{k}r}\right\rangle $). For such states,
then, we have an estimated ratio%
\[
\frac{\left\langle \left\vert \delta q_{\mathbf{k}r}(t_{f})\right\vert
\right\rangle _{\mathrm{eq}}}{\Delta q_{\mathbf{k}r}(t_{f})}\sim4ka_{f}%
\sqrt{\left\langle \hat{n}_{\mathbf{k}r}\right\rangle _{f}+1/2}\int_{t_{i}%
}^{t_{f}}dt\ \frac{1}{a^{2}}\sqrt{\left\langle \hat{n}_{\mathbf{k}%
r}\right\rangle +1/2}\ .
\]
It then follows that if (instead of (\ref{condn}))%
\begin{equation}
\frac{1}{k}\lesssim4a_{f}\sqrt{\left\langle \hat{n}_{\mathbf{k}r}\right\rangle
_{f}+1/2}\int_{t_{i}}^{t_{f}}dt\ \frac{1}{a^{2}}\sqrt{\left\langle \hat
{n}_{\mathbf{k}r}\right\rangle +1/2}\ , \label{n-condn}%
\end{equation}
or equally if (instead of (\ref{condn2}))%
\begin{equation}
4a_{f}\sqrt{a_{f}\left\langle \hat{H}_{\mathbf{k}r}\right\rangle _{f}}%
\int_{t_{i}}^{t_{f}}dt\ \frac{1}{a^{2}}\sqrt{a\left\langle \hat{H}%
_{\mathbf{k}r}\right\rangle }\gtrsim1\ , \label{n-condn2}%
\end{equation}
then%
\begin{equation}
\frac{\left\langle \left\vert \delta q_{\mathbf{k}r}(t_{f})\right\vert
\right\rangle _{\mathrm{eq}}}{\Delta q_{\mathbf{k}r}(t_{f})}\gtrsim1\ .
\end{equation}
From this we may reasonably deduce that relaxation, or at least significant
relaxation, is likely to occur (except of course for special states with very
simple velocity fields).

Unlike our proof of relaxation suppression for modes satisfying (\ref{condn}),
this is not a rigorous result. (It is roughly analogous to saying, in
classical kinetic theory, that significant relaxation to thermal equilibrium
is likely to occur, over timescales of order the mean free time, if the mean
magnitude of momentum transferred in molecular collisions is comparable to the
width of the equilibrium momentum distribution.) To delineate the precise
behaviour in this region requires further study, perhaps through numerical simulations.

To avoid potential misunderstandings, we should emphasise that relaxation
might of course be suppressed for special quantum states violating the
freezing inequality (\ref{condn}) (in particular, states with an especially
simple de Broglie velocity field). However, one should bear in mind that we
are concerned with the evolution of quantum nonequilibrium in our actual
universe, which is known to have had a complex and violent past history. Thus,
for example, in a standard radiation-dominated phase, special states with no
entanglement at any time are of no interest: we are concerned with states that
are likely to have actually occurred. In seeking a general criterion for the
freezing of early nonequilibrium, it is then of no use to point to special
quantum states exhibiting particularly simple velocity fields.\footnote{A
notable exception is the inflationary vacuum, which is in fact an example of a
state that is non-entangled (across modes), with a very simple velocity field,
and which is widely believed to have existed in the past; see section 10.1.}
In contrast, the freezing inequality (\ref{condn}) is a natural constraint on
quantum states in general, providing a realistic pointer to where
nonequilibrium might be found in our actual universe. And violation of
(\ref{condn}) is, as we have argued in this section, likely to imply
relaxation or at least significant relaxation.

\section{Time Evolution of $\left\langle \hat{n}_{\mathbf{k}r}\right\rangle $
for a Free Field}

For a given expansion history $a=a(t)$ on $[t_{i},t_{f}]$, the freezing
inequality (\ref{condn}) depends on the time evolution of $\left\langle
\hat{n}_{\mathbf{k}r}\right\rangle _{t}$ on $[t_{i},t_{f}]$ (or,
(\ref{condn2}) depends on $\left\langle \hat{H}_{\mathbf{k}r}\right\rangle
_{t}$). To make precise predictions, then, we require a specific cosmological
model, and an explicit expression for $\left\langle \hat{n}_{\mathbf{k}%
r}\right\rangle _{t}$ as a function of time $t$. We leave such detailed
studies for future work. Here, we give a method for calculating $\left\langle
\hat{n}_{\mathbf{k}r}\right\rangle _{t}$ (and $\left\langle \hat
{H}_{\mathbf{k}r}\right\rangle _{t}$) for an arbitrary pure quantum state.
This method might prove useful.

The mean energy%
\[
W_{\mathbf{k}r}\equiv\left\langle \hat{H}_{\mathbf{k}r}\right\rangle
=(\left\langle \hat{n}_{\mathbf{k}r}\right\rangle +1/2)(k/a)
\]
in the mode $\mathbf{k}r$ evolves in time according to $dW_{\mathbf{k}%
r}/dt=\left\langle \partial\hat{H}_{\mathbf{k}r}/\partial t\right\rangle $,
which implies (using $\dot{a}=Ha$)%
\begin{equation}
\frac{dW_{\mathbf{k}r}}{dt}=-3HW_{\mathbf{k}r}+4HU_{\mathbf{k}r}\ ,
\end{equation}
where $U_{\mathbf{k}r}\equiv\left\langle \frac{1}{2}a^{3}\omega^{2}\hat
{q}_{\mathbf{k}r}^{2}\right\rangle $ is the mean potential energy. (For an
interacting field, as discussed in section 5.2, $dW_{\mathbf{k}r}/dt$ would
contain additional terms from $-i\left\langle [\hat{H}_{\mathbf{k}r},\hat
{H}_{I}]\right\rangle $.) The rate of change of $\left\langle \hat
{n}_{\mathbf{k}r}\right\rangle =(a/k)W_{\mathbf{k}r}-1/2$ is then given by%
\begin{equation}
\frac{d\left\langle \hat{n}_{\mathbf{k}r}\right\rangle }{dt}%
=-2(Ha/k)(K_{\mathbf{k}r}-U_{\mathbf{k}r})\ ,
\end{equation}
where $K_{\mathbf{k}r}\equiv\left\langle \hat{\pi}_{\mathbf{k}r}^{2}%
/2a^{3}\right\rangle $ is the mean kinetic energy.

To solve for $W_{\mathbf{k}r}(t)=K_{\mathbf{k}r}(t)+U_{\mathbf{k}r}(t)$, and
hence the required function $\left\langle \hat{n}_{\mathbf{k}r}\right\rangle
_{t}$, one may write first-order (linear) differential equations for
$K_{\mathbf{k}r}$, $U_{\mathbf{k}r}$ and for the quantity $\chi_{\mathbf{k}%
r}\equiv\frac{1}{2}\left\langle \hat{q}_{\mathbf{k}r}\hat{\pi}_{\mathbf{k}%
r}+\hat{\pi}_{\mathbf{k}r}\hat{q}_{\mathbf{k}r}\right\rangle $. Using
$d\left\langle \hat{\Omega}\right\rangle /dt=-i\left\langle [\hat{\Omega}%
,\hat{H}]\right\rangle +\left\langle \partial\hat{\Omega}/\partial
t\right\rangle $, it is readily shown that%
\begin{equation}
\frac{dK_{\mathbf{k}r}}{dt}=-3HK_{\mathbf{k}r}-\omega^{2}\chi_{\mathbf{k}%
r}\ ,\ \ \ \ \ \frac{dU_{\mathbf{k}r}}{dt}=HU_{\mathbf{k}r}+\omega^{2}%
\chi_{\mathbf{k}r}\ ,\ \ \ \ \ \frac{d\chi_{\mathbf{k}r}}{dt}=2(K_{\mathbf{k}%
r}-U_{\mathbf{k}r})\ . \label{odes}%
\end{equation}
If $H=\dot{a}/a$ and $\omega=k/a$ are known functions of time, then given
values of $K_{\mathbf{k}r}$, $U_{\mathbf{k}r}$, $\chi_{\mathbf{k}r}$ at any
one time (say $t_{i}$ or $t_{f}$) --- where these values are determined by the
wave functional $\Psi$ at that time\footnote{Of course, initial values for
$K_{\mathbf{k}r}$, $U_{\mathbf{k}r}$, $\chi_{\mathbf{k}r}$ cannot be chosen
completely arbitrarily. They are subject to constraints, such as
$a^{3}K+U/(ak^{2})+\chi=\frac{1}{2}\left\langle (\hat{\pi}+\hat{q}%
)^{2}\right\rangle \geq0$ (or $ak^{2}K+U/(a^{3})+\omega^{2}\chi\geq0$).} ---
the equations (\ref{odes}) determine $K_{\mathbf{k}r}$, $U_{\mathbf{k}r}$,
$\chi_{\mathbf{k}r}$ at all times, yielding $W_{\mathbf{k}r}(t)=K_{\mathbf{k}%
r}(t)+U_{\mathbf{k}r}(t)$ as well as the required function $\left\langle
\hat{n}_{\mathbf{k}r}\right\rangle _{t}=a(t)W_{\mathbf{k}r}(t)/k-1/2$.

Introducing the vector $X=(K_{\mathbf{k}r},U_{\mathbf{k}r},\chi_{\mathbf{k}%
r})^{\mathrm{T}}$, the equations (\ref{odes}) take the form $dX/dt=AX$, where
$A$ is the time-dependent matrix%
\begin{equation}
A=%
\begin{pmatrix}
-3\dot{a}/a & 0 & -k^{2}/a^{2}\\
0 & \dot{a}/a & k^{2}/a^{2}\\
2 & -2 & 0
\end{pmatrix}
\ .
\end{equation}
For interesting forms of $a$, such as $a\propto t^{1/2}$, it seems likely that
these equations will have to be solved numerically.

It would be interesting to study this system of equations, and to establish
the conditions under which solutions for $\left\langle \hat{n}_{\mathbf{k}%
r}\right\rangle _{t}$ (or $W_{\mathbf{k}r}(t)=\left\langle \hat{H}%
_{\mathbf{k}r}\right\rangle _{t}$) satisfy the freezing inequality
(\ref{condn}) (or (\ref{condn2})). We leave this for future work.

\section{Approximate Solutions for $\left\langle \hat{n}_{\mathbf{k}%
r}\right\rangle _{t}$ Satisfying the Freezing Inequality}

However, it is important to show first of all that solutions for $\left\langle
\hat{n}_{\mathbf{k}r}\right\rangle _{t}$ satisfying (\ref{condn}) can exist
for some values of $k$. Here, we construct approximate solutions of
(\ref{odes}) valid in the long-wavelength limit $k\rightarrow0$, that satisfy
(\ref{condn}) for appropriate initial conditions and time intervals. The
conditions of validity are probably too restrictive for useful application to
realistic cosmological scenarios, and we give these solutions here only to
show that solutions satisfying (\ref{condn}) are indeed possible.

We consider a radiation-dominated expansion, for which $a\propto t^{1/2}$ and
$H=1/2t$. Dropping the indices $\mathbf{k}r$, we find approximate solutions to
(\ref{odes}) satisfying (for appropriate values of $k$)%
\begin{equation}
\omega^{2}|\chi|<<HK,\ HU \label{approx}%
\end{equation}
(where $K,\ U$ are non-negative), or%
\begin{equation}
\frac{k^{2}t_{i}}{a_{i}^{2}}|\chi|<<K,\ U \label{approx1}%
\end{equation}
(where $t_{i}/a_{i}^{2}=t_{f}/a_{f}^{2}=t/a^{2}$). We then have the simple
solutions%
\begin{equation}
K=K_{i}(t/t_{i})^{-3/2}\ ,\ \ \ \ \ U=U_{i}(t/t_{i})^{1/2} \label{KU}%
\end{equation}
(that is, $K\propto1/a^{3}$ and $U\propto a$), and%
\begin{equation}
\chi=\chi_{i}+4K_{i}t_{i}\left(  1-(t/t_{i})^{-1/2}\right)  +\frac{4}{3}%
U_{i}t_{i}\left(  1-(t/t_{i})^{3/2}\right)  \ . \label{chi}%
\end{equation}

Note that, for these solutions, the quantities $\left\langle \hat
{q}_{\mathbf{k}r}^{2}\right\rangle =2U_{\mathbf{k}r}/(k^{2}a)$ and
$\left\langle \hat{\pi}_{\mathbf{k}r}^{2}\right\rangle =2a^{3}K_{\mathbf{k}r}$
are time independent.

We need to show the consistency of the solutions (\ref{KU}) and (\ref{chi})
with the assumed approximation (\ref{approx1}). This may be done if $k$ is
appropriately small. Specifically, writing%
\[
\chi=\chi_{i}+4K_{i}t_{i}+\frac{4}{3}U_{i}t_{i}-4Kt-\frac{4}{3}Ut\ ,
\]
we have (since $Kt$ and $Ut$ respectively decrease and increase with time)%
\[
|\chi|\leq|\chi_{i}|+8K_{i}t_{i}+\frac{8}{3}U_{f}t_{f}\equiv D\ .
\]
If we assume that%
\begin{equation}
k^{2}<<\frac{a_{i}^{2}}{t_{i}^{2}}\frac{t_{i}}{D}\min\left\{  K_{f}%
,U_{i}\right\}  \label{consis}%
\end{equation}
(where $(t_{i}/D)\min\left\{  K_{f},U_{i}\right\}  $ is dimensionless), we
then have%
\[
\frac{k^{2}t_{i}}{a_{i}^{2}}|\chi|\leq\frac{k^{2}t_{i}}{a_{i}^{2}}%
D<<\min\left\{  K_{f},U_{i}\right\}  \leq K,\ U
\]
(since $K$ and $U$ respectively decrease and increase), and so the
approximation condition (\ref{approx1}) is indeed satisfied.

For $k$ satisfying (\ref{consis}), we then have the approximate solutions
(\ref{KU}) for $K$ and $U$. We wish to show explicitly that, for these
solutions, there are values of $k$ that satisfy the freezing inequality
(\ref{condn}) (or (\ref{condn2})).

To show this, for simplicity we first choose initial conditions with
$K_{i}<<U_{i}$. Since $K$ decreases with time, we then have $\min\left\{
K_{f},U_{i}\right\}  =K_{f}$ and (from (\ref{consis})) the solutions
(\ref{KU}) are valid if%
\begin{equation}
k^{2}<<\frac{a_{i}^{2}}{t_{i}}\frac{K_{f}}{D} \label{consis2}%
\end{equation}
(where $a_{i}^{2}/t_{i}=a_{f}^{2}/t_{f}=a^{2}/t$). Further, since $K$
decreases and $U$ increases with time, $K_{i}<<U_{i}$ implies that $K<<U$ for
all $t\geq t_{i}$. Thus we have $\left\langle \hat{H}\right\rangle _{t}\approx
U(t)$ (where we continue to suppress the indices $\mathbf{k}r$), or (using
(\ref{KU}))%
\begin{equation}
\left\langle \hat{H}\right\rangle _{t}\approx U_{i}(t/t_{i})^{1/2}\ .
\label{Hamt1}%
\end{equation}
Inserting this into the freezing inequality (\ref{condn2}), and using
$a=a_{f}(t/t_{f})^{1/2}$ and $H_{f}^{-1}=2t_{f}$, and taking $t_{i}/t_{f}<<1$,
we obtain%
\begin{equation}
U_{i}<\frac{1}{4}\frac{a_{i}}{a_{f}}\frac{1}{H_{f}^{-1}}\ . \label{Ucon}%
\end{equation}
Since $\left\langle \hat{H}\right\rangle _{i}\approx U_{i}$ we have (restoring
indices $\mathbf{k}r$) the freezing inequality%
\begin{equation}
\left\langle \hat{H}_{\mathbf{k}r}\right\rangle _{i}<\frac{1}{4}\frac{a_{i}%
}{a_{f}}\frac{1}{H_{f}^{-1}}=\frac{1}{4}\left(  \frac{a_{i}}{a_{f}}\right)
^{3}\frac{1}{H_{i}^{-1}}\ . \label{Hcon}%
\end{equation}
(Note that, since for the above solution $\left\langle \hat{H}_{\mathbf{k}%
r}\right\rangle $ increases with time, the general result (\ref{hbar}) also
applies, with $\left\langle \hat{H}_{\mathbf{k}r}\right\rangle _{i}%
=\left\langle \hat{H}_{\mathbf{k}r}\right\rangle _{\min}<1/8H_{f}^{-1}$. This
is consistent with (\ref{Hcon}), since we have assumed $t_{i}/t_{f}<<1$ which
implies $a_{i}/a_{f}<<1$.)

Thus, for a given mode $\mathbf{k}r$ satisfying (\ref{consis2}), if
$\left\langle \hat{H}_{\mathbf{k}r}\right\rangle _{i}$ is sufficiently small
(satisfying (\ref{Hcon})), then relaxation will be suppressed and initial
nonequilibrium (if it exists) will be frozen. And it is indeed always possible
to choose $\left\langle \hat{H}_{\mathbf{k}r}\right\rangle _{i}$ so as to
satisfy (\ref{Hcon}), provided $k$ is sufficiently small. For the only general
constraint on $\left\langle \hat{H}_{\mathbf{k}r}\right\rangle _{i}$ is%
\[
\left\langle \hat{H}_{\mathbf{k}r}\right\rangle _{i}=(\left\langle \hat
{n}_{\mathbf{k}r}\right\rangle _{i}+\frac{1}{2})\frac{k}{a_{i}}\geq\frac{1}%
{2}\frac{k}{a_{i}}\ ,
\]
so it is possible to satisfy (\ref{Hcon}) if%
\[
k<\frac{1}{2}\frac{a_{i}}{a_{f}}\frac{a_{i}}{H_{f}^{-1}}%
\]
or%
\[
\lambda_{\mathrm{phys}}=a_{f}\lambda>4\pi\left(  \frac{a_{f}}{a_{i}}\right)
^{2}H_{f}^{-1}>>H_{f}^{-1}\ .
\]

If instead we choose initial conditions with $K_{i}>>U_{i}$, we will have
$K>>U$ only for as long as $t_{i}/t_{f}$ is not much smaller than $1$. Over
this limited time, we have%
\begin{equation}
\left\langle \hat{H}\right\rangle _{t}\approx K(t)=K_{i}(t/t_{i})^{-3/2}%
=K_{f}(t/t_{f})^{-3/2}\ . \label{Hamt2}%
\end{equation}
Since $K_{f}>>U_{f}>U_{i}$, we now have $\min\left\{  K_{f},U_{i}\right\}
=U_{i}$ and the solution is valid if%
\begin{equation}
k^{2}<<\frac{a_{i}^{2}}{t_{i}}\frac{U_{i}}{D}\ . \label{consis3}%
\end{equation}
Inserting (\ref{Hamt2}) into the freezing inequality (\ref{condn2}), and
assuming that $t_{i}/t_{f}$ is small compared to $1$ (but not so small as to
invalidate the approximation $K>>U$), we obtain the freezing inequality%
\begin{equation}
\left\langle \hat{H}_{\mathbf{k}r}\right\rangle _{i}\approx K_{i}<\frac{1}%
{4}\frac{1}{H_{i}^{-1}}\ . \label{Hcon2}%
\end{equation}
Again using $\left\langle \hat{H}_{\mathbf{k}r}\right\rangle _{i}%
\geq(1/2)(k/a_{i})$, we now find that it is possible to satisfy (\ref{Hcon2})
if%
\[
k<\frac{1}{2}\frac{a_{i}}{H_{i}^{-1}}=\frac{1}{2}\left(  \frac{a_{f}}{a_{i}%
}\right)  ^{2}\frac{a_{i}}{H_{f}^{-1}}%
\]
or%
\[
\lambda_{\mathrm{phys}}=a_{f}\lambda>4\pi\frac{a_{i}}{a_{f}}H_{f}^{-1}\ .
\]
By assumption, $a_{i}/a_{f}$ is not much smaller than $1$, so we still have
$\lambda_{\mathrm{phys}}\gtrsim H_{f}^{-1}$. (In any case, it follows from
(\ref{consis3}) that the solution is valid only if $a_{f}\lambda>H_{f}^{-1}$.
For we have%
\[
D\geq8K_{i}t_{i}+\frac{8}{3}U_{f}t_{f}>8K_{i}t_{i}+\frac{8}{3}U_{i}%
t_{i}\approx8K_{i}t_{i}\ ,
\]
so that (\ref{consis3}) gives%
\[
k^{2}<<\frac{a_{i}^{2}}{t_{i}}\frac{U_{i}}{D}\lesssim\frac{a_{i}^{2}}%
{t_{i}^{2}}\frac{U_{i}}{8K_{i}}<<\left(  \frac{a_{i}}{t_{i}}\right)  ^{2}%
\]
or $a_{f}\lambda>>(a_{i}/a_{f})H_{f}^{-1}$. Since $a_{i}/a_{f}$ is not much
smaller than $1$, we indeed have $a_{f}\lambda>H_{f}^{-1}$.)

It is therefore certainly possible to have solutions for $\left\langle \hat
{n}_{\mathbf{k}r}\right\rangle _{t}$ with nonequilibrium freezing for
long-wavelength modes, $a_{f}\lambda>H_{f}^{-1}$ (or $a_{f}\lambda>>H_{f}%
^{-1}$).

\section{Possible Consequences of Early Nonequilibrium Freezing}

The freezing inequality (\ref{condn}) (or (\ref{condn2})) makes it possible,
for the first time, to make quantitative predictions for nonequilibrium
deviations from quantum theory, if we are given a specific cosmological model.
The potential consequences are many, and much remains to be done to develop
them. Here, we restrict ourselves to a preliminary sketch of some possible
nonequilibrium effects, in particular: corrections to inflationary predictions
for the CMB, non-inflationary super-Hubble field correlations, and relic
nonequilibrium particles. We hope to develop further details elsewhere, in the
context of specific (and realistic) cosmological models.

As we saw in section 6, for a radiation-dominated expansion (\ref{condn})
implies the general lower bound (\ref{lam-con}) on the physical wavelength
$\lambda_{\mathrm{phys}}(t_{f})$ --- the wavelength of what might be termed
`relic nonequilibrium field modes' --- at the final time $t_{f}$. In terms of
the ambient temperature $T$, where $T\propto1/a\propto t^{-1/2}$, the lower
bound may be written as%
\begin{equation}
\lambda_{\mathrm{phys}}(t_{f})>4\pi H_{f}^{-1}\ln(T_{i}/T_{f})\ .
\label{lam-con2}%
\end{equation}
As we have discussed, this lower bound will in practice be not more than two
or three orders of magnitude larger than the Hubble radius $H_{f}^{-1}$ at
time $t_{f}$.

Note that, to satisfy the freezing inequality, the bound (\ref{lam-con2}) is a
necessary but not sufficient condition. A detailed understanding of where
nonequilibrium freezing can occur requires, as discussed in section 5.1, a
calculation of the time evolution of the mean occupation numbers $\left\langle
\hat{n}_{\mathbf{k}r}\right\rangle _{\alpha}$ for the pure subensembles (with
wave functionals $\Psi_{\alpha}$) contained in the early mixed state, to find
out which --- if any --- of these subensembles satisfy (\ref{condn}). This is
a matter for future work. Here, we consider only the necessary condition
(\ref{lam-con2}), which provides a pointer to where residual nonequilibrium
\textit{could} be found (pending the said more complete analysis). In
particular, (\ref{lam-con2}) suggests that one should look for nonequilibrium
above a specific critical wavelength.

\subsection{Corrections to Inflationary Predictions for the CMB}

In inflationary cosmology, the universe undergoes a period of exponential
expansion, $a(t)\propto e^{Ht}$, driven by the energy density of an
approximately homogeneous scalar or inflaton field $\phi$, where quantum
fluctuations in $\phi$ seed the primordial curvature perturbations that are
later imprinted as temperature anisotropies in the CMB \cite{LL00}.

To a first approximation, inflation predicts that modes of the inflaton field
will have a quantum variance%
\begin{equation}
\left\langle |\phi_{\mathbf{k}}|^{2}\right\rangle _{\mathrm{QT}}=\frac
{V}{2(2\pi)^{3}}\frac{H^{2}}{k^{3}} \label{BD}%
\end{equation}
and a scale-invariant power spectrum%
\begin{equation}
\mathcal{P}_{\phi}^{\mathrm{QT}}(k)\equiv\frac{4\pi k^{3}}{V}\left\langle
\left\vert \phi_{\mathbf{k}}\right\vert ^{2}\right\rangle _{\mathrm{QT}}%
=\frac{H^{2}}{4\pi^{2}}\ , \label{HZ}%
\end{equation}
where $\left\langle \left\vert \phi_{\mathbf{k}}\right\vert ^{2}\right\rangle
_{\mathrm{QT}}$ is obtained from the Bunch-Davies vacuum in de Sitter space,
for $\lambda_{\mathrm{phys}}>>H^{-1}$. In the slow-roll limit ($\dot
{H}\rightarrow0$), this results in a scale-invariant spectrum, $\mathcal{P}%
_{\mathcal{R}}^{\mathrm{QT}}(k)=\mathrm{const}.$, for the primordial curvature
perturbation $\mathcal{R}_{\mathbf{k}}$, in approximate agreement with what is
observed in the CMB \cite{5YrWMAP}.

Now, quantum nonequilibrium in the early Bunch-Davies vacuum generally implies
deviations from (\ref{BD}). It has been shown \cite{AV07,AV08} that if
(microscopic) quantum nonequilibrium exists at the onset of inflation, then
instead of relaxing it will be preserved during the inflationary phase, and
furthermore it will be transferred to macroscopic lengthscales by the
expansion of physical wavelengths $\lambda_{\mathrm{phys}}\propto a(t)\propto
e^{Ht}$. Specifically, for each mode $\mathbf{k}$, explicit calculation shows
that the width of the evolving nonequilibrium distribution remains in a
constant ratio with the width of the equilibrium distribution. (This is
essentially because the vacuum state has the special property of being
non-entangled across modes, so that the de Broglie-Bohm trajectories decompose
into independent one-dimensional motions. See ref. \cite{AV08}.) If we write
the nonequilibrium variance as%
\begin{equation}
\left\langle |\phi_{\mathbf{k}}|^{2}\right\rangle =\left\langle |\phi
_{\mathbf{k}}|^{2}\right\rangle _{\mathrm{QT}}\xi(k)
\end{equation}
(where equilibrium corresponds of course to $\xi(k)=1$ for all $k$), the power
spectrum for $\mathcal{R}_{\mathbf{k}}$ is then just the quantum result
multiplied by the `nonequilibrium factor' $\xi(k)$: that is, $\mathcal{P}%
_{\mathcal{R}}(k)=\mathcal{P}_{\mathcal{R}}^{\mathrm{QT}}(k)\xi(k)$.

Thus, quantum nonequilibrium at the beginning of inflation will generally
break the scale invariance of $\mathcal{P}_{\mathcal{R}}(k)$. As discussed in
detail elsewhere \cite{AV08}, measurements of the angular power spectrum for
the CMB may be used (in the context of inflation) to set bounds on $\xi(k)$.

Given these results, the next step is to try to predict some features of the
function $\xi(k)$. This requires a constraint on the form of nonequilibrium at
the onset of inflation.

One possible strategy is to consider a pre-inflationary era, and to derive
constraints on residual nonequilibrium from that era. If we take the
pre-inflationary era to be radiation-dominated ($a\propto t^{1/2}$), the lower
bound (\ref{lam-con2}) shows that nonequilibrium (for whatever fields may be
present in that era) can survive only for sufficiently large, super-Hubble
wavelengths. Since $\lambda_{\mathrm{phys}}\propto t^{1/2}$ and $H^{-1}\propto
t$, at sufficiently early times all physical wavelengths will in fact be
super-Hubble ($\lambda_{\mathrm{phys}}>H^{-1}$), raising the possibility of
nonequilibrium freezing for the corresponding modes (if the freezing
inequality (\ref{condn})\ is satisfied). During the subsequent inflationary
phase, $H^{-1}$ is (approximately) constant, and relevant cosmological
fluctuations originate from inside $H^{-1}$. Some of these fluctuating modes
could be out of equilibrium only if they evolved from modes that were outside
the Hubble radius in the pre-inflationary phase.

Thus, in order to obtain nonequilibrium corrections to inflationary
predictions for the CMB, arising from an earlier pre-inflationary era, some of
the pre-inflationary nonequilibrium modes must enter the Hubble radius, and
they must avoid complete relaxation by the time inflation begins. Because
pre-inflationary modes with larger values of $\lambda$ enter the Hubble radius
later, they are presumably less likely to relax before inflation begins, in
which case residual nonequilibrium will be possible only for $\lambda$ larger
than some infra-red cutoff $\lambda_{\mathrm{c}}$. (For further discussion,
see ref. \cite{AV08}.)

We hope that future work, based on a specific pre-inflationary model, will
provide a prediction for $\lambda_{\mathrm{c}}$, as well as some indication of
the form of the nonequilibrium spectrum for $\lambda\gtrsim\lambda
_{\mathrm{c}}$. Note that $\xi(k)<1$ at wave number $k$ implies that the
nonequilibrium width of the corresponding inflaton mode is less than the
equilibrium width. One might reasonably expect this, in view of the hypothesis
that quantum noise arose from statistical relaxation processes in the very
early universe: it seems natural to assume that early nonequilibrium would
have a less-than-quantum dispersion, $\xi(k)<1$, as opposed to a
larger-than-quantum dispersion, $\xi(k)>1$ (though the latter is of course
possible in principle). Thus, a dip $\xi(k)<1$ in the power spectrum below
some critical wave number $k_{\mathrm{c}}=2\pi/\lambda_{\mathrm{c}}$ might be
naturally explained in terms of quantum nonequilibrium surviving from a very
early pre-inflationary era.

It has in fact been found that an infra-red cutoff in the primordial power
spectrum provides a slightly better fit to the 3-year WMAP data; however, the
improvement is not sufficient to justify introducing the additional cutoff
parameter in the model \cite{IRcutoff}.

\subsection{Super-Hubble Correlations without Inflation?}

As noted in the introduction, one motivation for assuming quantum
nonequilibrium at the big bang was that the resulting nonlocality at early
times could eliminate the cosmological horizon problem (which persists, as we
have mentioned, even in some inflationary models \cite{VT00}). One might also
ask if early quantum nonequilibrium could provide an alternative,
non-inflationary means of laying down primordial curvature perturbations at
super-Hubble lengthscales, in a standard Friedmann cosmology. Since we have
shown that nonequilibrium can remain frozen at super-Hubble scales, one may
ask if such nonequilibrium could generate appropriate super-Hubble
correlations without the need for an inflationary era.

The Bunch-Davies vacuum for a scalar field $\phi$, with variance given by
(\ref{BD}) (at long wavelengths), has the remarkable property that the
two-point correlation function%
\[
\left\langle \phi(\mathbf{x}_{1})\phi(\mathbf{x}_{2})\right\rangle =\frac
{1}{V}\int d^{3}\mathbf{k}\;e^{i\mathbf{k}\cdot(\mathbf{x}_{1}-\mathbf{x}%
_{2})}\left\langle |\phi_{\mathbf{k}}|^{2}\right\rangle
\]
is independent of distance $\left\vert \mathbf{x}_{1}-\mathbf{x}%
_{2}\right\vert $, as is readily verified for $\left\langle |\phi_{\mathbf{k}%
}|^{2}\right\rangle =\left\langle |\phi_{\mathbf{k}}|^{2}\right\rangle
_{\mathrm{QT}}\propto1/k^{3}$. As a first step, one may ask how this
inflationary quantum behaviour could be mimicked by a non-inflationary vacuum
in quantum nonequilibrium.

Consider a vacuum state whose quantum variance is $\left\langle |\phi
_{\mathbf{k}}|^{2}\right\rangle _{\mathrm{QT}}\propto k^{m_{\mathrm{QT}}}$ for
some fixed index $m_{\mathrm{QT}}$. Assuming that the quantum two-point
function decreases with distance, where $\left\langle \phi(\mathbf{x}_{1}%
)\phi(\mathbf{x}_{2})\right\rangle _{\mathrm{QT}}\propto\left\vert
\mathbf{x}_{1}-\mathbf{x}_{2}\right\vert ^{-(m_{\mathrm{QT}}+3)}$, we have
$m_{\mathrm{QT}}>-3$. Then consider the same vacuum state in quantum
nonequilibrium, with $\left\langle |\phi_{\mathbf{k}}|^{2}\right\rangle
=\left\langle |\phi_{\mathbf{k}}|^{2}\right\rangle _{\mathrm{QT}}\xi(k)$,
assuming that $\xi(k)\propto k^{\mu}$ for some fixed index $\mu$. To obtain a
nonequilibrium two-point function that is independent of distance, we require
$m_{\mathrm{QT}}+\mu=-3$, or $\mu<0$, so that (in this simple example) the
nonequilibrium function $\xi(k)$ must increase as $k\rightarrow0$.

As things stand, we are unable to say if such behaviour for $\xi(k)$ is likely
to emerge from any reasonable model. However, given the upper bound (\ref{+})
on the ratio $\left\langle \left\vert \delta q_{\mathbf{k}r}(t_{f})\right\vert
\right\rangle _{\mathrm{eq}}/\Delta q_{\mathbf{k}r}(t_{f})$, one could study
how the `degree of freezing' varies with $k$ (for example for $k\rightarrow
0$), where a high or low degree of freezing could be defined respectively as a
low or high value of the upper bound on the right-hand-side of (\ref{+}). For
a specific cosmological model, with some assumptions about initial conditions,
this could provide constraints on the behaviour of the function $\xi(k)$. The
results will obviously depend on how $\left\langle \hat{n}_{\mathbf{k}%
r}\right\rangle $ varies with $k$.

Finally, we note that a nonlocal model, based not on hidden variables or
quantum nonequilibrium but on the holographic principle, has been shown to
generate the required (approximately) scale-invariant perturbation spectrum at
super-Hubble scales \cite{MSC07}. Whether or not early quantum nonequilibrium
could reproduce such effects in a natural way remains to be seen.

\subsection{Relic Nonequilibrium Particles}

We saw in section 10.1 that relic nonequilibrium field modes could, in the
case of the inflaton, change the power spectrum for primordial curvature
perturbations, resulting in observable effects in the CMB. Thus, inflationary
cosmology provides a simple and definite means whereby early nonequilibrium
could yield observable consequences today. In this section, in contrast, we
shall attempt to outline some much more complicated and uncertain scenarios,
according to which relic nonequilibrium field modes (for some appropriate
field) might in some circumstances manifest as relic nonequilibrium particles
that could be detected today. Unfortunately, for these scenarios to be at all
plausible, some questionable assumptions have to be made, and at the time of
writing it is not clear if these scenarios can really work in practice.

There is of course no preferred definition of particle states in quantum field
theory on expanding space, except in the short-wavelength limit (where one
recovers the usual Minkowski definition) \cite{BD}. The very notion of
`particles' is in fact highly ambiguous for modes of frequency lower than the
typical inverse timescale over which the spacetime metric changes. In a
cosmological setting, this means that there is no generally useful definition
of quantum particle states at wavelengths larger than the Hubble radius. Thus,
if we consider relic nonequilibrium field modes from a radiation-dominated era
--- where the bound (\ref{lam-con2}) implies that such modes must have
super-Hubble wavelengths --- we must be careful not to interpret such modes
too naively in terms of (quantum) particle states. However, pending a more
precise treatment, one might reasonably assume that if such modes enter the
Hubble radius at later times, then they will manifest as
(approximately-defined) particle states in the usual sense.

One should also bear in mind that, generally speaking, excitations of
super-Hubble modes will not be produced by the local processes of particle
scattering and decay (which are not expected to be effective over lengthscales
larger than the instantaneous Hubble radius $H^{-1}$). However, such
excitations will of course be produced by the global effects of spatial expansion.

In order to maximise the chance of obtaining relic nonequilibrium particles
that could be detected in practice (in particular, with energies that are not
so low as to be completely out of range), we ought to try to minimise the
lower bound on the mode wavelength defined by (\ref{lam-con}) or
(\ref{lam-con2}). This can be done by choosing the final time $t_{f}$ to be as
small as possible, subject to the constraint that further relaxation may be
neglected for times later than the chosen value of $t_{f}$. Thus, one might
take $t_{f}$ to be the time $t_{\mathrm{dec}}$ at which the relevant particle
species decouples. For one might reasonably assume that relaxation may be
neglected (at all wavelengths) for $t>t_{\mathrm{dec}}$ --- if the quantum
states, defined post-decoherence, are such that the associated de Broglie
velocity field is sufficiently simple (as occurs, for example, for energy
eigenstates). For a super-Hubble mode at $t_{\mathrm{dec}}$ that becomes
sub-Hubble at later times, it is then conceivable that any nonequilibrium
present at the time $t_{\mathrm{dec}}$ could persist until much later. (A
proper discussion of this scenario would require an analysis of decoherence
before and after decoupling.)

If we make the above assumptions, the key question is then whether residual
nonequilibrium field modes can exist at the time $t_{f}=t_{\mathrm{dec}}$.
From (\ref{lam-con2}), this is possible if the modes have physical wavelength
(inserting the Boltzmann constant $k_{\mathrm{B}}$)\-%
\begin{equation}
\lambda_{\mathrm{phys}}(t_{\mathrm{dec}})>4\pi H_{\mathrm{dec}}^{-1}%
\ln(k_{\mathrm{B}}T_{i}/k_{\mathrm{B}}T_{\mathrm{dec}})\ , \label{lam-con3}%
\end{equation}
where $H_{\mathrm{dec}}^{-1}$ and $T_{\mathrm{dec}}$ are respectively the
Hubble radius and temperature at time $t_{\mathrm{dec}}$. We have
$\lambda_{\mathrm{phys}}(t_{\mathrm{dec}})=a_{\mathrm{dec}}\lambda$, where
$a_{\mathrm{dec}}=T_{0}/T_{\mathrm{dec}}$ (with $T_{0}\simeq2.7\ \mathrm{K}$
the temperature today). Assuming that decoupling occurs before the end of the
radiation-dominated phase, we also have $H_{\mathrm{dec}}^{-1}%
=2t_{\mathrm{dec}}$, where $t_{\mathrm{dec}}$ may be expressed in terms of
$T_{\mathrm{dec}}$ using the standard temperature-time relation%
\begin{equation}
t\sim(1\ \mathrm{s})\left(  \frac{1\ \mathrm{MeV}}{k_{\mathrm{B}}T}\right)
^{2}\ . \label{tT}%
\end{equation}
The lower bound (\ref{lam-con3}) then becomes (inserting the speed of light
$c$)%
\begin{equation}
\lambda\gtrsim8\pi c(1\ \mathrm{s})\left(  \frac{1\ \mathrm{MeV}%
}{k_{\mathrm{B}}T_{\mathrm{dec}}}\right)  \left(  \frac{1\ \mathrm{MeV}%
}{k_{\mathrm{B}}T_{0}}\right)  \ln\left(  \frac{k_{\mathrm{B}}T_{i}%
}{k_{\mathrm{B}}T_{\mathrm{dec}}}\right)  \,,
\end{equation}
or (with $c\simeq3\times10^{10}\ \mathrm{cm\,s}^{-1}$, $k_{\mathrm{B}}%
\simeq8.6\times10^{-5}\ \mathrm{eV\,K}^{-1}$, and $k_{\mathrm{B}}T_{0}%
\simeq2.3\times10^{-4}\ \mathrm{eV}$)%
\begin{equation}
\lambda\gtrsim(3.3\times10^{21}\ \mathrm{cm})\left(  \frac{1\ \mathrm{MeV}%
}{k_{\mathrm{B}}T_{\mathrm{dec}}}\right)  \ln\left(  \frac{k_{\mathrm{B}}%
T_{i}}{k_{\mathrm{B}}T_{\mathrm{dec}}}\right)  \,. \label{lb}%
\end{equation}
This provides a lower bound on the wavelength $\lambda$ \textit{today}, at
which nonequilibrium could be found.

The freezing inequality (\ref{condn}), and the resulting lower bound
(\ref{lb}), have been derived in this paper for massless scalar fields only.
One certainly expects to find comparable results for more general massless
boson fields, such as the electromagnetic field. For fermions, however, a
separate analysis is required. There are different approaches to the
pilot-wave theory of fermions, and the details of nonequilibrium freezing may
depend on which model is adopted. One might try to derive a fermionic analogue
of the freezing inequality using, for example, the Dirac sea pilot-wave model
\cite{CS07}. Pending such extensions of our analysis, here we assume that the
lower bound (\ref{lb}) applies (at least approximately) to fermions as well,
provided they are effectively massless at the temperature $T_{\mathrm{dec}}$
(that is, of mass $m<<k_{\mathrm{B}}T_{\mathrm{dec}}/c^{2}$).

With this understanding, let us now apply the approximate result (\ref{lb}) to
various particle species, both bosonic and fermionic. For definiteness, we
first consider a standard Friedmann cosmology with no inflationary period,
taking our initial conditions at the Planck era $k_{\mathrm{B}}T_{i}\sim
k_{\mathrm{B}}T_{\mathrm{P}}\sim10^{19}\ \mathrm{GeV}$. (An alternative
possibility, of nonequilibrium relic particles arising from the decay of the
inflaton, is considered below.)

Photons decouple from matter at $k_{\mathrm{B}}(T_{\mathrm{dec}})_{\gamma}%
\sim0.3\ \mathrm{eV}$. From (\ref{lb}) we then have a lower bound%
\begin{equation}
\lambda_{\gamma}\gtrsim0.7\times10^{30}\ \mathrm{cm}\ ,
\end{equation}
which exceeds the Hubble radius today, $H_{0}^{-1}\simeq10^{28}\ \mathrm{cm}$.
If instead we consider neutrinos, which decouple at $k_{\mathrm{B}%
}(T_{\mathrm{dec}})_{\nu}\sim1\ \mathrm{MeV}$, we have%
\begin{equation}
\lambda_{\nu}\gtrsim1.7\times10^{23}\ \mathrm{cm}\simeq5.5\times
10^{4}\ \mathrm{pc}%
\end{equation}
(or $\sim10^{5}$ light years). Residual nonequilibrium for relic neutrinos
could plausibly exist today only at such tiny energies. Unfortunately, this is
of course far outside any realistic range of detection. (Note, again, the
implicit assumption being made, that if nonequilibrium super-Hubble modes at
$t_{\mathrm{dec}}$ enter the Hubble radius at $t>t_{\mathrm{dec}}$, they will
manifest as nonequilibrium particle states.)

The situation improves drastically, however, if one considers particles that
decouple soon after the Planck era. Gravitons, for example, are expected to
decouple at a temperature $(T_{\mathrm{dec}})_{g}\lesssim T_{\mathrm{P}}$.
Writing%
\[
k_{\mathrm{B}}(T_{\mathrm{dec}})_{g}\equiv x_{g}(k_{\mathrm{B}}T_{\mathrm{P}%
})\simeq x_{g}(10^{19}\ \mathrm{GeV})\ ,
\]
where $x_{g}\lesssim1$, we obtain%
\begin{equation}
\lambda_{g}\gtrsim(0.3\ \mathrm{cm})(1/x_{g})\ln\left(  1/x_{g}\right)  \,.
\label{lbg}%
\end{equation}
This might be compared with the range of wavelengths expected for a (thermal)
relic graviton background, whose temperature today is estimated to be
$(T_{0})_{g}\sim1\ \mathrm{K}$ \cite{Wein}. At this temperature, the spectral
energy density of a Planck distribution peaks at the wavelength $\lambda
_{\max}(1\ \mathrm{K})\simeq0.3\ \mathrm{cm}$.

There may also exist other particles that decouple not too long after the
Planck era, and that (unlike the graviton) are unstable, eventually producing
decay products that could be more easily detected today. A natural candidate,
arising out of current supersymmetric theories of high-energy physics, is the
unstable gravitino $\tilde{G}$, which has been estimated to decouple at a
temperature \cite{FY02}%
\[
k_{\mathrm{B}}(T_{\mathrm{dec}})_{\tilde{G}}\equiv x_{\tilde{G}}%
(k_{\mathrm{B}}T_{\mathrm{P}})\approx(1\ \mathrm{TeV})\left(  \frac{g_{\ast}%
}{230}\right)  ^{1/2}\left(  \frac{m_{\tilde{G}}}{10\ \mathrm{keV}}\right)
^{2}\left(  \frac{1\ \mathrm{TeV}}{m_{gl}}\right)  ^{2}\ ,
\]
where $g_{\ast}$ is the number of spin degrees of freedom (for the effectively
massless particles) at the temperature $(T_{\mathrm{dec}})_{\tilde{G}}$,
$m_{gl}$ is the gluino mass, and $m_{\tilde{G}}$ is the gravitino mass. This
provides us with an estimate for the lower bound in the case of gravitinos,%
\begin{equation}
\lambda_{\tilde{G}}\gtrsim(0.3\ \mathrm{cm})(1/x_{\tilde{G}})\ln\left(
1/x_{\tilde{G}}\right)  \,. \label{lbG}%
\end{equation}
For the purposes of illustration, if we take $\left(  g_{\ast}/230\right)
^{1/2}\sim1$ and $\left(  1\ \mathrm{TeV/}m_{gl}\right)  ^{2}\sim1$, then%
\[
x_{\tilde{G}}\approx\left(  \frac{m_{\tilde{G}}}{10^{3}\ \mathrm{GeV}}\right)
^{2}\ .
\]
If, for example, $m_{\tilde{G}}\approx100\ \mathrm{GeV}$, then $x_{\tilde{G}%
}\approx10^{-2}$ and (\ref{lbG}) yields $\lambda_{\tilde{G}}\gtrsim
140\ \mathrm{cm}$. This corresponds to energies that are rather low, but
perhaps accessible.

If the gravitino is not the lightest supersymmetric particle, then it will
indeed be unstable. For large $m_{\tilde{G}}$, the total decay rate is
estimated to be \cite{NY06} $\Gamma_{\tilde{G}}=(193/48)(m_{\tilde{G}}%
^{3}/M_{\mathrm{P}}^{2})$, where $M_{\mathrm{P}}\simeq1.2\times10^{19}%
\ \mathrm{GeV}$ is the Planck mass. The time $(t_{\mathrm{decay}})_{\tilde{G}%
}$ at which the gravitino decays is of order the lifetime $1/\Gamma_{\tilde
{G}}$. Using (\ref{tT}), the corresponding temperature is%
\[
k_{\mathrm{B}}(T_{\mathrm{decay}})_{\tilde{G}}\sim(m_{\tilde{G}}%
/1\ \mathrm{GeV})^{3/2}\ \mathrm{eV}\ .
\]
For example, again for the case $m_{\tilde{G}}\approx100\ \mathrm{GeV}$, the
relic gravitinos decay when $k_{\mathrm{B}}(T_{\mathrm{decay}})_{\tilde{G}%
}\sim1\ \mathrm{keV}$. This is prior to photon decoupling, so that any
(potentially nonequilibrium) photons produced by the decaying gravitinos would
interact strongly with matter and quickly relax to quantum equilibrium. To
obtain gravitino decay after photon decoupling, we would need $k_{\mathrm{B}%
}(T_{\mathrm{decay}})_{\tilde{G}}\lesssim k_{\mathrm{B}}(T_{\mathrm{dec}%
})_{\gamma}\sim0.3\ \mathrm{eV}$, or $m_{\tilde{G}}\lesssim0.5\ \mathrm{GeV}$.
For such small gravitino masses, however, decoupling occurs at (roughly)%
\[
(T_{\mathrm{dec}})_{\tilde{G}}=x_{\tilde{G}}T_{\mathrm{P}}\approx\left(
m_{\tilde{G}}/10^{3}\ \mathrm{GeV}\right)  ^{2}T_{\mathrm{P}}\lesssim
10^{-7}T_{\mathrm{P}}%
\]
and (\ref{lbG}) (with $x_{\tilde{G}}\lesssim10^{-7}$) yields the much larger
lower bound $\lambda_{\tilde{G}}\gtrsim10^{7}\ \mathrm{cm}$. Thus, it may
prove more promising to consider other decay products (that decouple prior to
gravitino decay but for larger gravitino masses). These could in turn decay
into photons at later times, or they might be detected directly.

There are of course strong constraints on the presence of gravitinos in
cosmological models, in particular from the abundance of light elements
emerging from big-bang nucleosynthesis and from limits on dark matter
abundance. These constraints have been extensively studied --- see, for
example, ref. \cite{Moroi} --- and the subject is an active area of current
research. Our hope is that an acceptable and compelling scenario will
eventually be found, satisfying the standard cosmological constraints and at
the same time allowing the possibility of relic nonequilibrium surviving in
particles that could be detected today. To develop such a scenario in detail
is a topic for future work.

So far in this section, we have assumed a standard (non-inflationary)
Friedmann expansion, with initial nonequilibrium at around the Planck era. An
alternative scenario is obtained if we consider relic nonequilibrium particles
in the context of inflationary cosmology. If inflation did occur, the density
of any relic particles (nonequilibrium or otherwise) from a pre-inflationary
era will of course be so diluted as to be completely undetectable today.
However, one may consider relic particles that were created at the end of
inflation, by the decay of the inflaton field itself.

As discussed in section 10.1, during inflation the inflaton field does not
relax to quantum equilibrium, and in fact the exponential expansion of space
transfers any initial nonequilibrium from microscopic to macroscopic
lengthscales. The inflaton field, then, is a prime candidate for a carrier of
primordial quantum nonequilibrium. As well as manifesting as statistical
anomalies in the CMB, such nonequilibrium in the inflaton field could manifest
as nonequilibrium in its decay products, where in standard inflationary
scenarios inflaton decay is in fact the source of the matter and radiation
present in our universe today.

The process of `preheating' is driven by the homogeneous and essentially
classical part of the inflaton field (that is, by the $k=0$ mode)
\cite{BTW06}. Here, the inflaton is treated as a classical external field,
acting on other (quantum) fields which become excited by parametric resonance.
Because of the classicality of the relevant part of the inflaton field, this
process is unlikely to result in a transference of nonequilibrium from the
inflaton to the created particles.

During `reheating', however, perturbative decay of the inflaton can occur, and
one may reasonably expect nonequilibrium in the inflaton field to be
transferred to its decay products. This possibility opens up a large field of
investigation. Here, again, we restrict ourselves to making some preliminary remarks.

The perturbative decay of the inflaton occurs through local field-theoretical
interactions, so one expects the decay products to have physical wavelengths
no greater than the instantaneous Hubble radius. Taking the lower bound
(\ref{lam-con2}) as a guide (even though it was derived for a
radiation-dominated phase), we then expect that the decay products will come
into existence already violating the freezing inequality. Subsequent
relaxation might then be avoided (possibly) only if the particles are created
at a temperature below their decoupling temperature. Once again, the gravitino
suggests itself as a possible candidate. Gravitinos can in fact be copiously
produced by inflaton decay \cite{Phidec} (and could even make up a significant
component of dark matter \cite{Tak07}). If the gravitinos are unstable, again,
one could try to detect (say) photons produced by their decay at later times.

The possible realisation of this scenario depends of course on uncertain
features of high-energy particle physics and of inflationary models. As
before, one may hope that a scenario will eventually be found, satisfying the
constraints of particle physics and cosmology, and at the same time allowing
the possibility of relic nonequilibrium surviving in particles that could be
detected today.

We close this section with some general remarks.

First, we note that particle decay (for example for the gravitino) is likely
to result in some relaxation and erasure of any quantum nonequilibrium that
may have existed in the parent particles. However, one hopes that the erasure
will not be complete and that some nonequilibrium will still be present in the
decay products. It would be useful to study this, in pilot-wave models of
specific decay processes.

Second, once suitable candidates for nonequilibrium relic particles have been
identified, one must consider how best to test them for violations of the Born
rule. For photons, a particularly simple test involves searching for anomalous
polarisation probabilities, or deviations from Malus' law (where such
deviations reflect the nonequilibrium breakdown of expectation additivity for
non-commuting quantum observables in a two-state system) \cite{AVSig,PV06}.

Third, for a given species of relic particle in the universe today, even if
there exist pure subensembles with significant residual nonequilibrium, in
practice it might be difficult for us to locate those subensembles and perform
experiments with them. In particular, if a given detector registers particles
belonging to different subensembles, without distinguishing between them, it
is possible that even if nonequilibrium is present in the individual
subensembles it will not be visible in the data.

\section{Conclusion}

The hypothesis of quantum nonequilibrium at the big bang has been shown to
have a number of observable consequences. Our main result is the freezing
inequality (\ref{condn}). For cosmological field modes satisfying
(\ref{condn}), initial nonequilibrium will be `frozen' at later times. This
result may be applied to specific cosmological models, yielding predictions
whose verification could constitute evidence for quantum nonequilibrium in our
universe. For a radiation-dominated expansion, (\ref{condn}) implies the
general lower bound (\ref{lam-con}) on the wavelength of relic nonequilibrium
field modes.

The detailed study of quantum nonequilibrium freezing, for realistic
cosmological models, is left for future work. A useful first step might be to
study the system of equations (\ref{odes}), and to delineate the general
conditions under which the time evolution of a (mean) mode occupation number
$\left\langle \hat{n}_{\mathbf{k}r}\right\rangle _{t}$ can satisfy the
freezing inequality (\ref{condn}). Crucially, future work will need to study
the statistical distribution of wave functionals for a realistic mixed state
on expanding space, the goal being to identify subensembles satisfying
(\ref{condn}). For these subensembles, quantum nonequilibrium is expected to
be frozen over the relevant time period, resulting in definite predictions
that might be tested today.

\textbf{Acknowledgements.} This work was partly supported by grant RFP1-06-13A
from The Foundational Questions Institute (fqxi.org). For their hospitality, I
am grateful to Carlo Rovelli and Marc Knecht at the Centre de Physique
Th\'{e}orique (Luminy), to Susan and Steffen Kyhl in Cassis, and to Jonathan
Halliwell at Imperial College London.


\begin{thebibliography}{99}                                                                                               %


\bibitem {deB28}L. de Broglie, in: \textit{\'{E}lectrons et Photons: Rapports
et Discussions du Cinqui\`{e}me Conseil de Physique}, ed. J. Bordet
(Gauthier-Villars, Paris, 1928). [English translation in ref. \cite{BV}.]

\bibitem {BV}G. Bacciagaluppi and A. Valentini, \textit{Quantum Theory at the
Crossroads: Reconsidering the 1927 Solvay Conference} (Cambridge University
Press, forthcoming) [quant-ph/0609184].

\bibitem {B52}D. Bohm, Phys. Rev. \textbf{85}, 166; 180 (1952).

\bibitem {AV91a}A. Valentini, Phys. Lett. A \textbf{156}, 5 (1991).

\bibitem {AV91b}A. Valentini, Phys. Lett. A \textbf{158}, 1 (1991).

\bibitem {AV92}A. Valentini, PhD thesis, International School for Advanced
Studies, Trieste, Italy (1992) [http://www.sissa.it/ap/PhD/Theses/valentini.pdf].

\bibitem {AV96}A. Valentini, in: \textit{Bohmian Mechanics and Quantum Theory:
an Appraisal}, eds. J. T. Cushing \textit{et al}. (Kluwer, Dordrecht, 1996).

\bibitem {AV01}A. Valentini, in: \textit{Chance in Physics: Foundations and
Perspectives}, eds. J. Bricmont \textit{et al}. (Springer, Berlin, 2001) [quant-ph/0104067].

\bibitem {AV02a}A. Valentini, Phys. Lett. A \textbf{297}, 273 (2002).

\bibitem {AV02b}A. Valentini, in: \textit{Non-Locality and Modality}, eds. T.
Placek and J. Butterfield (Kluwer, Dordrecht, 2002) [quant-ph/0112151].

\bibitem {AVPr02}A. Valentini, Pramana -- J. Phys. \textbf{59}, 269 (2002).

\bibitem {AVSig}A. Valentini, Phys. Lett. A \textbf{332}, 187 (2004).

\bibitem {AVBHs}A. Valentini, hep-th/0407032.

\bibitem {PV06}P. Pearle and A. Valentini, in: \textit{Encyclopaedia of
Mathematical Physics}, eds. J.-P. Fran\c{c}oise \textit{et al}. (Elsevier,
North-Holland, 2006) [quant-ph/0506115].

\bibitem {AV07}A. Valentini, J. Phys. A: Math. Theor. \textbf{40}, 3285 (2007).

\bibitem {AVSim}A. Valentini, in: \textit{Einstein, Relativity and Absolute
Simultaneity}, eds. W. L. Craig and Q. Smith (Routledge, London, 2008) [quant-ph/0504011].

\bibitem {VW05}A. Valentini and H. Westman, Proc. Roy. Soc. Lond. A
\textbf{461}, 253 (2005).

\bibitem {Sky}P. Fernstr\"{o}m, J. Johansson and A. Skyman, Chalmers
University of Technology unpublished report.

\bibitem {VT00}T. Vachaspati and M. Trodden, Phys. Rev. D \textbf{61}, 023502 (2000).

\bibitem {AV08}A. Valentini, Inflationary cosmology as a probe of primordial
quantum mechanics, arXiv, to appear.

\bibitem {BH84}D. Bohm and B. J. Hiley, Found. Phys. \textbf{14}, 255 (1984).

\bibitem {BHK87}D. Bohm, B. J. Hiley and P. N. Kaloyerou, Phys. Rep.
\textbf{144}, 321 (1987).

\bibitem {Holl93}P. R. Holland, \textit{The Quantum Theory of Motion: an
Account of the de Broglie-Bohm Causal Interpretation of Quantum Mechanics}
(Cambridge University Press, Cambridge, 1993).

\bibitem {BandH}D. Bohm and B. J. Hiley, \textit{The Undivided Universe: an
Ontological Interpretation of Quantum Theory} (Routledge, London, 1993).

\bibitem {Kal94}P. N. Kaloyerou, Phys. Rep. \textbf{244}, 287 (1994).

\bibitem {StruyPR}W. Struyve, Phys. Rep. (to appear) [arXiv:0707.3685].

\bibitem {Shank}R. Shankar, \textit{Principles of Quantum Mechanics} (Plenum
Press, New York, 1980).

\bibitem {Pad93}T. Padmanabhan, \textit{Structure Formation in the Universe}
(Cambridge University Press, Cambridge, 1993).

\bibitem {LL00}A. R. Liddle and D. H. Lyth, \textit{Cosmological Inflation and
Large-Scale Structure} (Cambridge University Press, Cambridge, 2000).

\bibitem {BH96}D. Bohm and B. J. Hiley, Found. Phys. \textbf{26}, 823 (1996).

\bibitem {Gold}S. Goldstein, J. L. Lebowitz, R. Tumulka and N. Zangh\`{\i}, J.
Stat. Phys. \textbf{125}, 1193 (2006).

\bibitem {Tum}R. Tumulka and N. Zangh\`{\i}, J. Math. Phys. \textbf{46},
112104 (2005).

\bibitem {5YrWMAP}G. Hinshaw \textit{et al}., arXiv:0803.0732 [astro-ph]; E.
Komatsu \textit{et al}., arXiv:0803.0547 [astro-ph].

\bibitem {IRcutoff}D. N. Spergel \textit{et al}., Astrophys. J. Supp.
\textbf{170}, 377 (2007).

\bibitem {MSC07}J. Magueijo, L. Smolin and C. R. Contaldi, Class. Quantum
Grav. \textbf{24}, 3691 (2007).

\bibitem {BD}N. D. Birrell and P. C. W. Davies, \textit{Quantum Fields in
Curved Space} (Cambridge University Press, Cambridge, 1982).

\bibitem {CS07}S. Colin and W. Struyve, J. Phys. A: Math. Theor. \textbf{40},
7309 (2007).

\bibitem {Wein}S. Weinberg, \textit{Gravitation and Cosmology: Principles and
Applications of the General Theory of Relativity} (John Wiley \& Sons, New
York, 1972).

\bibitem {FY02}M. Fujii and T. Yanagida, Phys. Lett. B \textbf{549}, 273 (2002).

\bibitem {NY06}S. Nakamura and M. Yamaguchi, Phys. Lett. B \textbf{638}, 389 (2006).

\bibitem {Moroi}T. Moroi, PhD thesis, Tohoku University, Sendai, Japan (1995) [hep-ph/9503210].

\bibitem {BTW06}B. A. Bassett, S. Tsujikawa and D. Wands, Rev. Mod. Phys.
\textbf{78}, 537 (2006).

\bibitem {Phidec}M. Endo, K. Hamaguchi and F. Takahashi, Phys. Rev. Lett.
\textbf{96}, 211301 (2006); M. Kawasaki, F. Takahashi and T. T. Yanagida,
Phys. Rev. D \textbf{74}, 043519 (2006); M. Endo, F. Takahashi and T. T.
Yanagida, Phys. Rev. D \textbf{76}, 083509 (2007).

\bibitem {Tak07}F. Takahashi, arXiv:0705.0579 [hep-ph].
\end{thebibliography}
\end{document}